\documentclass[11pt]{article}
\usepackage{putex}
\usepackage{graphicx}

\begin{document}
\preprint{
COLO-HEP-583 \\ PUPT-2457}

\institution{CU}{${}^1$Department of Physics, 390 UCB, University of Colorado, Boulder, CO 80309, USA}
\institution{PU}{${}^2$Joseph Henry Laboratories, Princeton University, Princeton, NJ 08544, USA}

\institution{CCTP}{${}^3$Crete Center for Theoretical Physics, University of
Crete, 71003 Heraklion, Greece}

\title{Minding the gap in ${\cal N}=4$ Super-Yang-Mills}

\authors{Oliver DeWolfe,${}^\CU$ Steven S.~Gubser,${}^\PU$ and Christopher Rosen${}^\CCTP$}

\abstract{We analyze  fermionic response in the geometry holographically dual to zero-temperature ${\cal N}=4$ Super-Yang-Mills theory with two equal nonvanishing chemical potentials, which is characterized by a singular horizon and zero ground state entropy. We show that  fermionic fluctuations are completely stable within a gap in energy around a Fermi surface singularity, beyond which non-Fermi liquid behavior returns. This gap disappears abruptly once the final charge is turned on, and is associated to a discontinuity in the corresponding chemical potential. We also show that the singular near-horizon geometry lifts to a smooth $AdS_3 \times \mathbb{R}^3$, and interpret the gap as a region where the quasiparticle momentum is spacelike in six dimensions due to the momentum component in the Kaluza-Klein direction, corresponding to the final charge.
}

\date{December 2013}

\maketitle

\section{Introduction and summary}

\subsection{Fermionic response in gauge/gravity systems}

Strongly coupled systems are by their nature difficult to study theoretically, yet they describe a number of vital and interesting phenomena in diverse areas of physics. To better understand such systems, it is valuable to develop alternate theoretical tools in which the physics is presented in a dual description using weakly-coupled variables. One such framework is the gauge/gravity correspondence, in which strongly-coupled field theories with a large number of degrees of freedom may be realized holographically as smooth gravity backgrounds in a  higher dimension.

Field theory states of nonzero temperature and/or density are realized in the correspondence as geometries including a black brane horizon, with equilibrium properties determined by  black hole thermodynamics.
We will be focused on zero temperature, nonzero density systems, which correspond to extremal black branes. Generically the event horizon for such an extremal brane is smooth, and due to the relationship between horizon area and entropy, such geometries are dual to zero-temperature field theory states with a nonzero, macroscopic entropy density. This large entropy is somewhat problematic for physical applications, as well as difficult to understand; explaining and potentially circumventing this feature is an important task.

To characterize the properties of these systems beyond basic thermodynamics, one may study linear response. Here we will be concerned with the response of the system to probe fermionic operators; one may identify Fermi surface singularities and study the spectrum of nearby fluctuations.
This has been studied substantially from the ``bottom-up" perspective, where the gravity background is not derived directly from string theory or supergravity; for this and other studies of nonzero-density systems using the gauge/gravity correspondence see \cite{Lee:2008xf,Liu:2009dm, Cubrovic:2009ye,Faulkner:2009wj} and many other papers
 \cite{Berkooz:2006wc}-
\cite{Davison:2013uha}; recent reviews appear in \cite{Hartnoll:2011fn, Iqbal:2011ae}.
Depending on parameters, these systems may manifest Fermi liquid, non-Fermi liquid, or marginal Fermi liquid behaviors.  In a Fermi liquid, quasiparticles become arbitrarily well defined as one approaches the Fermi surface at zero temperature.  In a non-Fermi liquid, this does not happen; instead the width of excitations remains comparable to or parametrically greater than their energies.  Marginal Fermi liquids lie on the border between the two classes. In the ``strange metals" arising in high-$T_c$ cuprates \cite{Varma, Anderson} and in heavy fermion systems \cite{Gegenwart},  Fermi surfaces can be identified from photoemission experiments, but the associated gapless excitations are not long-lived, suggesting an association with non-Fermi liquids.

It is natural to wish to embed this kind of calculation in supergravity/string theory; such a ``top-down" construction has the  advantages of explicit knowledge of the field theory dual and the operator map, as well as putting to rest concerns that there might be something unphysical about the bottom-up construction. 
Such a top-down study of fermionic response was investigated in \cite{DeWolfe:2011aa} for the four- and five-dimensional geometries dual to the  maximally supersymmetric conformal theories in three and four dimensions, respectively ABJM theory and ${\cal N}=4$ Super-Yang-Mills (SYM) theory, and Fermi surfaces were found. The ${\cal N}=4$ SYM case  was studied in much more detail in \cite{DeWolfe:2012uv}. Earlier top-down studies of the gravitino sector, which did not find Fermi surface singularities, can be found in \cite{Gauntlett:2011mf, Belliard:2011qq, Gauntlett:2011wm}.

 ${\cal N}=4$ SYM is characterized by three independent chemical potentials. Extremal black holes\footnote{We will be careless about the distinction between black holes and branes; all our systems have planar horizons.} with all three chemical potentials nonzero have nonsingular horizons and the associated nonzero entropy at zero temperature; we will call these ``regular" cases.  These were exhaustively studied for the case of two equal chemical potentials in \cite{DeWolfe:2012uv}, where the Fermi surface behavior for all spin-1/2 fermionic modes not mixing with a gravitino was obtained. The results were all non-Fermi liquids, with a single case asymptotically approaching a marginal Fermi liquid; no ordinary Fermi liquids were in evidence.

It is desirable to understand the entropy that shows up in a generic zero-temperature background of ${\cal N}=4$ SYM. Moreover, it is interesting to consider limits where this zero-temperature entropy is absent. There is such a class of geometries in the ${\cal N}=4$ SYM family, where one of the chemical potentials vanishes; the simplest example is when the other two charges are equal, which we will call the ``two-charge black hole" (2QBH). The vanishing entropy for the extremal two-charge black hole is associated to a zero-area horizon, which is singular \cite{Gubser:2009qt}. Probe fermions were studied \cite{Gubser:2009qt} and the Dirac equation solved exactly in this background in \cite{Gubser:2012yb}. As with the bottom-up models, Fermi liquid, marginal Fermi liquid and non-Fermi liquid cases were all possible in these studies.
Other recent studies of the 2QBH include \cite{Alishahiha:2012ad, Kulaxizi:2012gy, Berkooz:2013tg, Davison:2013txa, Davison:2013uha}.

However, the full story is more subtle. The 2QBH background is a solution of the 5D gauged supergravity coming from the reduction of type IIB supergravity on $AdS_5 \times S^5$, and hence corresponds to a specific state in ${\cal N}=4$ super-Yang-Mills. However, the fermion actions studied in \cite{Gubser:2009qt,Gubser:2012yb} were not deduced directly from supergravity, but instead were postulated to take a simple form with a constant mass, making them bottom-up excitations in a top-down background.
In \cite{DeWolfe:2012uv}, the proper top-down fluctuation equations of the fermions of ${\cal N}=8$ supergravity were worked out, and they displayed a property not covered by the analysis of \cite{Gubser:2012yb}: the masses depended on the running scalar field, which diverges at the horizon/singularity. Such scalar couplings are common in supergravity, and one might imagine that if a zero-entropy extremal geometry always has a singularity, then it will be generic for fermionic fluctuation equations to have such divergences. Thus, it is possible that the true nature of fermionic response in such a background can only be uncovered by including these couplings, and one would like to study them in more detail. This process was begun in \cite{DeWolfe:2012uv}, and we continue it here.

\subsection{The gap in the two-charge black hole}

\begin{figure}
\begin{center}
\includegraphics[scale=1.0]{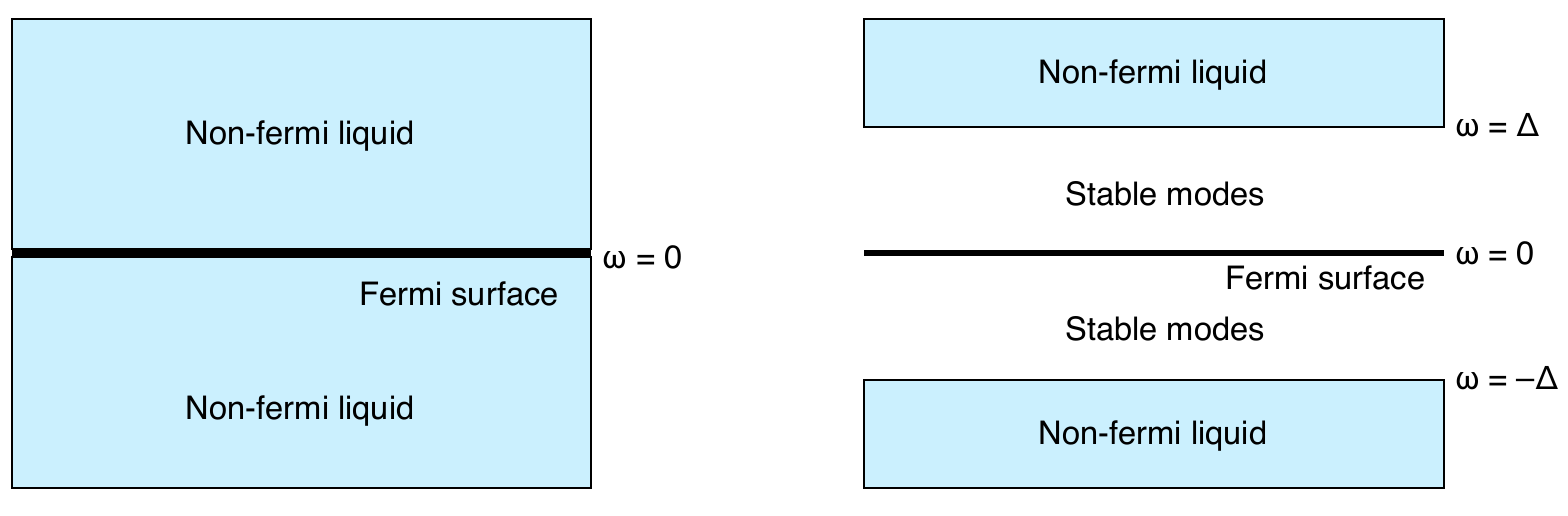}
\caption{Cartoon comparison between a ``regular" extremal black brane background (left) and the extremal two-charge black hole (right). In the former, fluctuations characteristic of a non-Fermi liquid surround the Fermi surface; in the latter, they are separated from the Fermi surface by a gap of size $2\Delta$ wherein the fluctuations are stable.
\label{GapSummaryFig}}
\end{center}
\end{figure}

To analyze the fermionic response in  finite density systems, we calculate retarded Green's functions for gauge-invariant fermionic operators; this is done on the gravity side by solving Dirac fluctuation equations with suitable infalling boundary conditions at the horizon. A pole in the Green's function at the Fermi energy (defined as $\omega = 0$) for some momentum $k = k_F$ defines a Fermi surface, and poles for nonzero $\omega$ describe the dispersion relation of nearby fluctuations. For ``regular" cases, the fluctuation energy is complex, $\omega \equiv \omega_* - i\Gamma$, including a real energy $\omega_*$ and a  width $\Gamma$ indicating the fluctuations are unstable. For Fermi liquids, the imaginary part of the Green's function vanishes faster than the real part as one approaches the Fermi surface, and  one has $\Gamma/\omega_* \to 0$.
In cases where the real and imaginary parts scale identically as one approaches the Fermi surface, on the other hand, one has $\Gamma/\omega_* \to$ constant, indicating
a non-Fermi liquid.
 
The essential novel feature of the extremal two-charge black hole is that there exists an energy scale $\Delta$ of order the chemical potential, which we call the ``gap", such that fluctuations within $\Delta$ of the Fermi surface have no decay width, and are thus precisely stable.\footnote{The gap we are discussing does not appear to be a manifestation of superconductivity, in that the $U(1)$ gauge fields under which the black hole is charged remain unbroken by any condensate, as does the final $U(1)$ gauge field under which the black hole is not charged.  It is possible however that the gap we observe is in some sense a precursor of superconductivity.} On the other side of the line $\omega  = \pm \Delta$, 
fluctuations reacquire a width  and behave analogously to excitations near the Fermi surface in regular cases.
We find that the real and imaginary parts of the Green's functions for ${\cal N}=4$ SYM scale identically in this region, characteristic of non-Fermi liquids, consistent with the results of \cite{DeWolfe:2012uv}; the constant that $\Gamma/\omega_*$ approaches, however, is zero, matching onto the stable region. This situation is summarized in cartoon form in figure~\ref{GapSummaryFig}.

A possible qualitative interpretation of this phenomenon runs as follows. In addition to whatever fermionic fluctuations we create, one postulates that there exists an additional sector with a large density of states. For a generic background in ${\cal N}=4$ SYM with all charges, this sector is not gapped and extends down to the Fermi surface, where its states are perceived as the nonzero ground state entropy. Fermionic fluctuations couple to this sector, and it mediates their ability to decay. In the special case of the two-charge black hole, however, this sector is gapped, and cannot be excited by energies less than $\Delta$ away from the Fermi surface. This removes the large entropy from zero temperature, and simultaneously removes the mechanism by which the fermionic fluctuations decay, rendering them stable in this energy range. We note that this implies that the fermionic fluctuations have no intrinsic self-interaction; one might speculate that this is a large-$N$ effect.

As one moves away from the 2QBH by adding even a tiny amount of the charge $Q_1$ that had been zero (corresponding to adding a charge density $\rho_1$), the gap disappears entirely. This behavior is associated to  another discontinuous phenomenon: the $\mu_1$ chemical potential jumps through zero from a positive to a negative constant as one varies from infinitessimal positive $\rho_1$ to infinitessimal negative $\rho_1$ (see figure~\ref{MuRhoFig}).
Such a discontinuity in the chemical potential can be characteristic of moving through a gap in the density of states, as the lack of states precludes the charge density from increasing even as the chemical potential is raised.

On the gravity side, the technical explanation for the gapped region is that the solution to the fluctuation equation becomes purely real. Associated to this, a true infalling boundary condition at the horizon is no longer possible; instead we impose a regularity condition, as is characteristic of Euclidean gauge/gravity calculations. We show this is the proper continuation of the infalling condition.

In the regular cases, the near-horizon and small-$\omega$ limits do not commute, necessitating the introduction of inner and outer regions which are then patched together \cite{Faulkner:2009wj}. The inner region has the geometry of $AdS_2 \times \mathbb{R}^3$ and determines the universal parts of the fermionic Green's function, including the decay width of quasiparticles.  For the 2QBH,  analogous phenomena occur at $|\omega| = \Delta$. The near-horizon geometry is singular in five dimensions and does not naively resemble anything as simple as $AdS_2$. However, we show that this geometry lifts to a smooth $AdS_3 \times \mathbb{R}^3$ in six dimensions, demonstrating that the singularity is of a ``good" type and the geometry is sensible. The Kaluza-Klein charge of the reduction is the charge $Q_1$ turned off in the 2QBH backgrounds, and the six-dimensional lift provides a new perspective on the gap $\Delta$, which can be understood as the minimum energy to turn a momentum vector with a fixed amount of compact momentum ($q_1$ charge) timelike. 
This suggests that a field theory explanation of the gap in the Green's functions of appropriately charged fermionic operators could be developed based on the emergence of a non-chiral Virasoro algebra in the infrared.

One can also derive consistency conditions on the charges and couplings of fermions that can be lifted to six dimensions. The discontinuity in the chemical potential $\mu_1$ is associated to a failure to commute of two different paths to
the extremal 2QBH in the black hole parameter space, and the form of the Dirac equation will depend on the path, leading to contradictory expressions, unless the fermions satisfy these conditions.
All the fermions of maximal gauged supergravity satisfy them, while a generic Dirac action will not,  once again suggesting that top-down considerations are an important ingredient, whereas a purely generic bottom-up fermion action may fail to properly capture the physics of these backgrounds.

Turning to specific examples, we analyze a number of distinct fermion species corresponding to specific operators of ${\cal N}=4$ SYM in the 2QBH, and numerically identify the dispersion relations through the stable region.
We find two classes of poles in the Green's function.  When there is a Fermi surface singularity at $\omega=0$, one may follow this singularity through the $\omega$-$k$ plane to the gap on either side. In addition, in some cases new pairs of poles nucleate at a nonzero value of $\omega$, and move towards the gap. These poles seem to be associated with the presence of an oscillatory region at $\omega = \Delta$ --- a region where the Green's function has log periodic behavior in $\omega$ and gapless excitations for a range of $k$, arising from an instability towards pair production in the near-horizon region of the geometry  \cite{Pioline:2005pf, Faulkner:2009wj}. These new poles generally reach the gap inside the oscillatory region, at which point they cease to exist, though in some cases one may miss it and survive.

The plan of this paper is as follows.  We review five-dimensional supergravity, the black brane solutions and their thermodynamics, and the fermionic fluctuation spectrum in section~\ref{GravitySec}.
The analysis of fermion fluctuations in regular extremal black holes is recapped in section~\ref{RegularSec}, before turning to the main subject of the paper in section~\ref{2QBHSec}, fermionic fluctuations in the 2QBH.
We apply these results to a number of fermionic operators of ${\cal N}=4$ Super-Yang-Mills in section~\ref{Nequals4Sec}. In section~\ref{6DSec} we present the lift of the 2QBH near-horizon geometry to $AdS_3 \times \mathbb{R}^3$ in six dimensions, and demonstrate how fermions are constrained by the lift. Some technical details are left for the appendices.

\section{Gravity dual of ${\cal N}=4$ Super-Yang-Mills at finite density}
\label{GravitySec}

\subsection{Supergravity and black brane backgrounds}

${\cal N}=4$ Super-Yang-Mills theory is the most symmetric avatar of four-dimensional gauge theory, and as such is often the gauge theory most amenable to study,  leading it to become 
 the simple harmonic oscillator of the twenty-first century study of quantum field theory. We are interested in states of finite density, associated to global conserved charges. The theory possesses an $SO(6)$ R-symmetry group; since this group is rank 3, there are three independent associated conserved quantities, and correspondingly one may turn on three distinct chemical potentials. Following \cite{DeWolfe:2012uv}, in what follows we will always simplify matters by taking two of the three chemical potentials equal; since the major distinct behaviors are associated with whether a given chemical potential is zero or not, this simplification seems still to capture the most interesting possibilities.

${\cal N}=4$ SYM with $SU(N)$ gauge group is holographically dual to type IIB string theory compactified on $AdS_5 \times S^5$ with $N$ units of self-dual five-form flux, with $SO(6)$ manifested as the isometry group of the five-sphere. The dynamics of the lowest-mass modes arising from the Kaluza-Klein reduction on $S^5$ are given by five-dimensional maximally supersymmetric (${\cal N}=8$) gauged supergravity, where $SO(6)$ becomes the gauge group. 
There exists a consistent truncation of the maximal gauged supergravity to a bosonic sector containing just the three gauge fields corresponding to the Cartan generators of $SO(6)$, along with the metric and two neutral scalars, the so-called STU model \cite{Behrndt:1998jd}. Simplifying matters as described above, we set two of the three gauge fields equal;  in this case only one scalar is sourced, and we are left with the effective gravity Lagrangian
\eqn{Lag}{
e^{-1} {\cal L} &=   R - {1 \over 2} (\partial \phi)^2 + {8 \over L^2} e^{\phi \over \sqrt{6}} + {4 \over L^2} e^{-2 \phi \over \sqrt{6}}
- e^{-4 \phi \over \sqrt{6}} f_{\mu\nu} f^{\mu\nu}  - 2
e^{2 \phi \over \sqrt{6}} F_{\mu\nu} F^{\mu\nu}  
- 2 \epsilon^{\mu\nu\rho\sigma\tau} f_{\mu\nu} F_{\rho\sigma} A_\tau \,.
}
Here $A_\mu$ is the gauge field associated to the two equal charges, and $a_\mu$ goes with the remaining one.

States of ${\cal N}=4$ SYM on $\mathbb{R}^{3,1}$ with nonzero temperature and chemical potentials are dual to geometries solving \eno{Lag} with the black-brane form
\eqn{Background}{
ds^2 &= e^{2A(r)} (- h(r) dt^2 + d \vec{x}^2) + {e^{2B(r)} \over h(r)} dr^2 \,, \cr
a_\mu dx^\mu &= \Phi_1(r)\, dt \,, \quad \quad A_\mu dx^\mu = \Phi_2(r)\, dt \,, \quad \quad \phi = \phi(r) \,,
}
with the functions
\eqn{TwoPlusOneSoln}{
A(r) &= \log {r \over L} + {1 \over 6} \log \left( 1 + {Q_1^2\over r^2} \right) + {1 \over 3} \log \left( 1 + {Q_2^2\over r^2} \right)\,, \cr
B(r) &= - \log {r \over L}- {1 \over 3} \log \left( 1 + {Q_1^2\over r^2} \right)- {2 \over 3} \log \left( 1 + {Q_2^2\over r^2} \right)\,, \cr
h(r) &= 1- { r^2(r_H^2 + Q_1^2)(r_H^2 + Q_2^2)^2 \over r_H^2 (r^2 + Q_1^2)(r^2 + Q_2^2)^2 }\,, \quad
\phi(r) =  -\sqrt{2\over 3} \log \left( 1 + {Q_1^2\over r^2} \right) +\sqrt{2\over 3} \log \left( 1 + {Q_2^2\over r^2} \right) \,, \cr
\Phi_1(r) &={Q_1 (r_H^2 + Q_2^2) \over 2 L r_H\sqrt{r_H^2 + Q_1^2}} \left(1-{r_H^2 + Q_1^2 \over r^2 + Q_1^2}   \right) \,, \quad
\Phi_2(r) ={Q_2 \sqrt{r_H^2 + Q_1^2} \over 2 L r_H} \left(1-{r_H^2 + Q_2^2 \over r^2 + Q_2^2} \right) \,.
}
The solutions are characterized by two charges $Q_1$, $Q_2$ and a horizon radius $r_H$ (the latter may be traded for a mass parameter). We refer to these as (2+1)-charge black holes ((2+1)QBHs).
The associated thermodynamics are
\eqn{}{
T &= {2 r_H^4 + Q_1^2 r_H^2 - Q_1^2 Q_2^2 \over 2 \pi L^2 r_H^2 \sqrt{r_H^2 + Q_1^2}} \,, \quad
\mu_1 = {Q_1 (r_H^2 + Q_2^2) \over L^2 r_H \sqrt{r_H^2 + Q_1^2}}\,, \quad
\mu_2 =  {\sqrt{2} Q_2 \sqrt{r_H^2 + Q_1^2} \over L^2 r_H} \,, \cr
s &= {1 \over 4 G L^3} (r_H^2 + Q_1^2)^{1/2} (r_H^2 + Q_2^2) \,, \quad \quad
\rho_1 = {Q_1 s \over 2 \pi r_H} \,, \quad \quad \rho_2 = {\sqrt{2} Q_2 s \over 2 \pi r_H} \,,
}
where $T$ and $s$ are the temperature and entropy density, $\mu_1$ and $\rho_1$ are the chemical potential and charge density for the single charge, and $\mu_2$ and $\rho_2$ are likewise for the two charges set equal. Geometries corresponding to zero temperature are extremal, with the constraint on the three parameters,
\eqn{ExtremalTwoPlusOne}{
2 r_H^4 + Q_1^2 r_H^2 - Q_1^2 Q_2^2  =0 \quad\quad \quad {\rm (extremal\ (2+1)QBH)}\,,
}
which has solutions for $|Q_2| > r_H$. For $Q_1,Q_2 \neq 0$, the extremal black holes are ``regular", possessing a double pole at the horizon in the  function $h(r)$ and no singularity. 
All these geometries  asymptotically approach (the boundary of) $AdS_5$ at $r \to \infty$,
\eqn{}{
e^{2A} \to {r^2 \over L^2} \,, \quad \quad e^{2B} \to {L^2 \over r^2} \,, \quad \quad h \to 1\,, \quad \quad \phi \to 0 \,, \quad \quad \Phi_i \to {\rm const} \,.
}
\begin{figure}
\begin{center}
\includegraphics[scale=0.5]{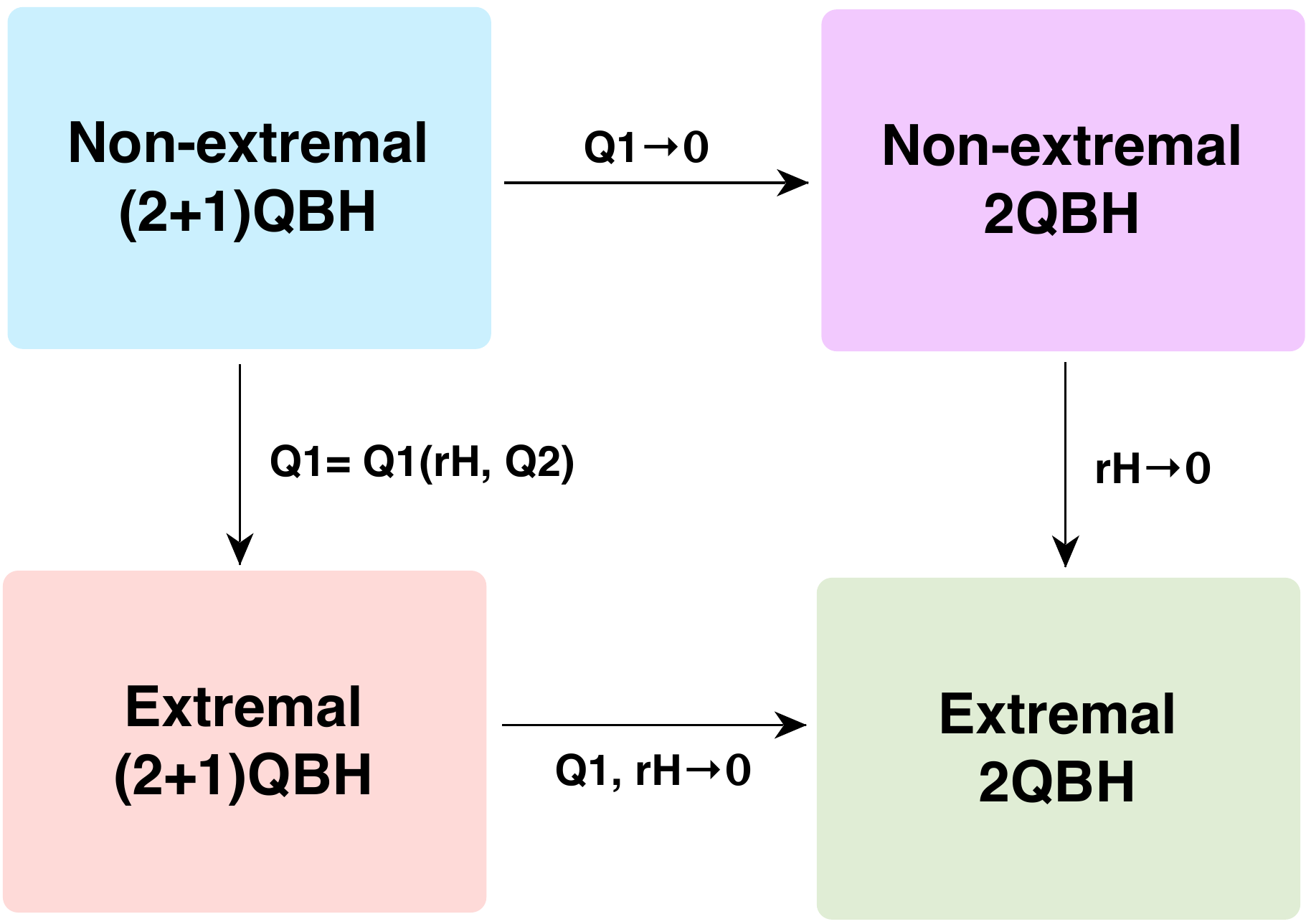}
\caption{ Two approaches to the extremal two-charge black hole.
\label{BHParamSpaceFig}}
\end{center}
\end{figure}
We will be focused on the special case where we take $Q_1 \to 0$ strictly; these are called the two-charge black holes, or 2QBHs. In this case $\Phi_1 = 0$. The 2QBH solution is 
\eqn{}{
A(r) &= \log {r \over L} + {1 \over 3} \log \left( 1 + {Q_2^2\over r^2} \right)\,, \quad
B(r) = - \log {r \over L}- {2 \over 3} \log \left( 1 + {Q_2^2\over r^2} \right)\,, \cr
h(r) &= 1- { (r_H^2 + Q_2^2)^2  \over (r^2 + Q_2^2)^2 }\,, \quad
\phi(r) =  \sqrt{2\over 3} \log \left( 1 + {Q_2^2\over r^2} \right) \,, \quad
\Phi_2(r) ={Q_2 \over2  L} \left(1-{r_H^2 + Q_2^2 \over r^2 + Q_2^2} \right) \,,
}
with thermodynamics
\eqn{}{
T = { r_H  \over  \pi L^2  } \,, \quad
\mu_2 =  {\sqrt{2} Q_2  \over L^2} \,, \quad
s = {r_H (r_H^2 + Q_2^2) \over 4 G L^3}   \,, \quad \quad
 \rho_2 = {\sqrt{2} Q_2 s \over 2 \pi r_H} \,,
}
and $\mu_1 = \rho_1 = 0$; we will sometimes refer to $\mu_2$ simply as $\mu$. Thus extremality occurs for
\eqn{}{
r_H = 0 \quad\quad\quad {\rm (extremal\ 2QBH)} \,.
}
The order of limits in reaching the extremal 2QBHs we are studying can be somewhat subtle.
We have so far described the 2QBH by first setting $Q_1 \to 0$, and then imposing extremality $r_H \to 0$. One may instead consider another path: first achieve extremality for (2+1)QBHs by imposing the relation \eno{ExtremalTwoPlusOne} between $r_H$, $Q_1$ and $Q_2$,
and then tune $Q_1$ to zero while maintaining extremality, which also requires adjusting $r_H \to 0$  (see figure~\ref{BHParamSpaceFig}). From the field theory point of view, this is tuning the charge density $\rho_1$ to zero while maintaining $T=0$.

 \begin{figure}
\begin{center}
\includegraphics[scale=0.7]{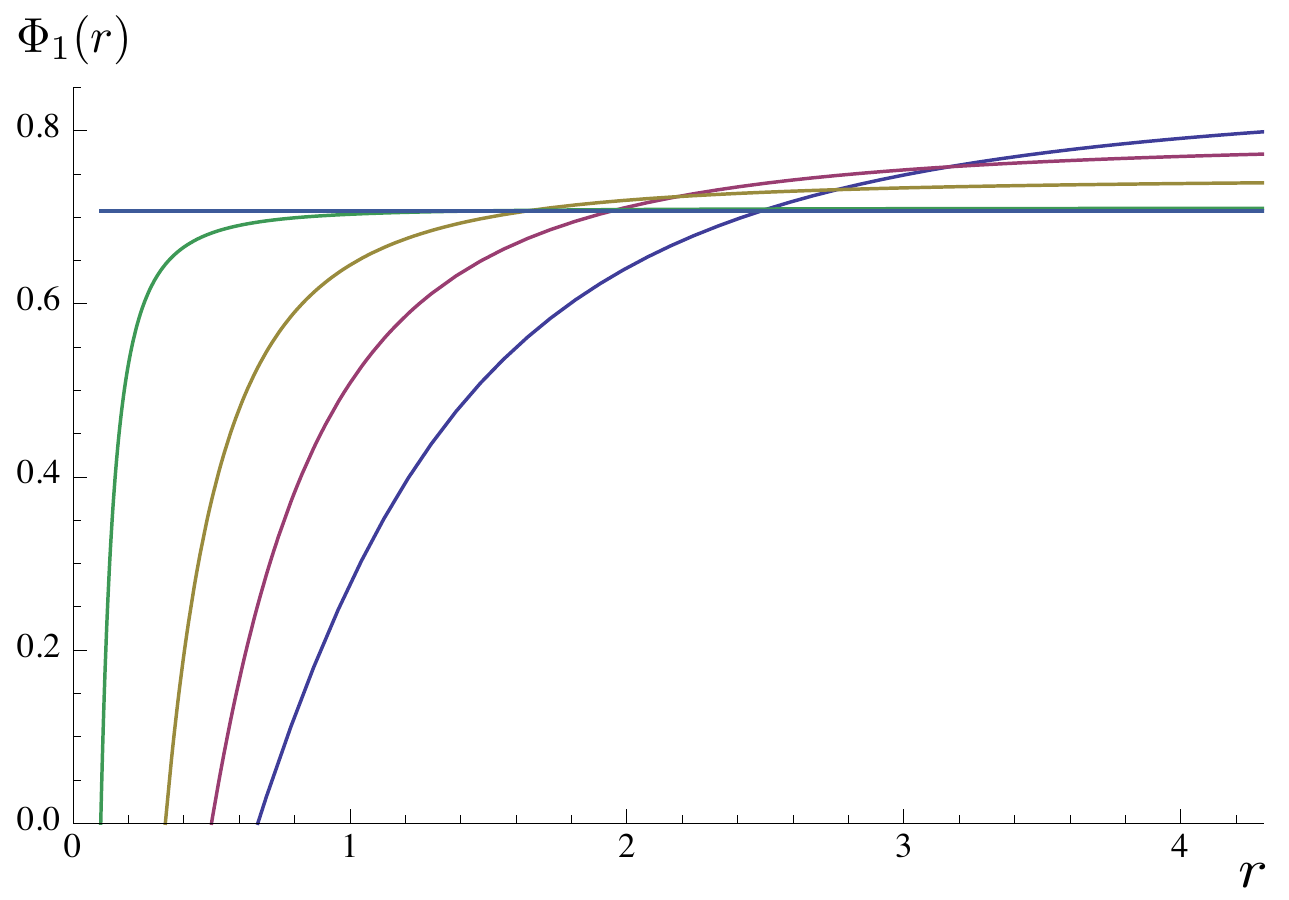}
\caption{Plots of $\Phi_1(r)$ for various $Q_1$ as $Q_1 \to 0$ with $T=0$ held fixed. For all nonzero $Q_1$ one has $\Phi_1(r_H) =0$; note $r_H$ moves left as $Q_1 \to 0$. For $Q_1 = 0$, $\Phi_1$ degenerates to a nonzero constant.
\label{Phi1Fig}}
\end{center}
\end{figure}

 As discussed in 
\cite{DeWolfe:2012uv}, these different limits are not precisely equivalent; the former implies
\eqn{}{
{Q_1 Q_2 \over r_H^2} \to 0 \quad \quad \quad \quad ({\rm extremal \; 2}) \,,
}
since $Q_1$ is taken strictly to zero first, while the latter instead gives
\eqn{QProduct}{
{Q_1 Q_2 \over r_H^2} \to \sqrt{2} \quad \quad \quad \quad ({\rm extremal \; 2+1}) \,.
}
This distinction affects only $\Phi_1(r)$. In the former case, it becomes zero strictly. In the latter case, however, one finds it approaching  the nonzero constant (see figure~\ref{Phi1Fig})
\eqn{Phi1Shift}{
\Phi_1 \to  {Q_2 \over \sqrt{2} L} \,.
}
While this is gauge-equivalent to $\Phi_1 = 0$, its form nonetheless affects the thermodynamics.
\begin{figure}
\begin{center}
\includegraphics[scale=0.7]{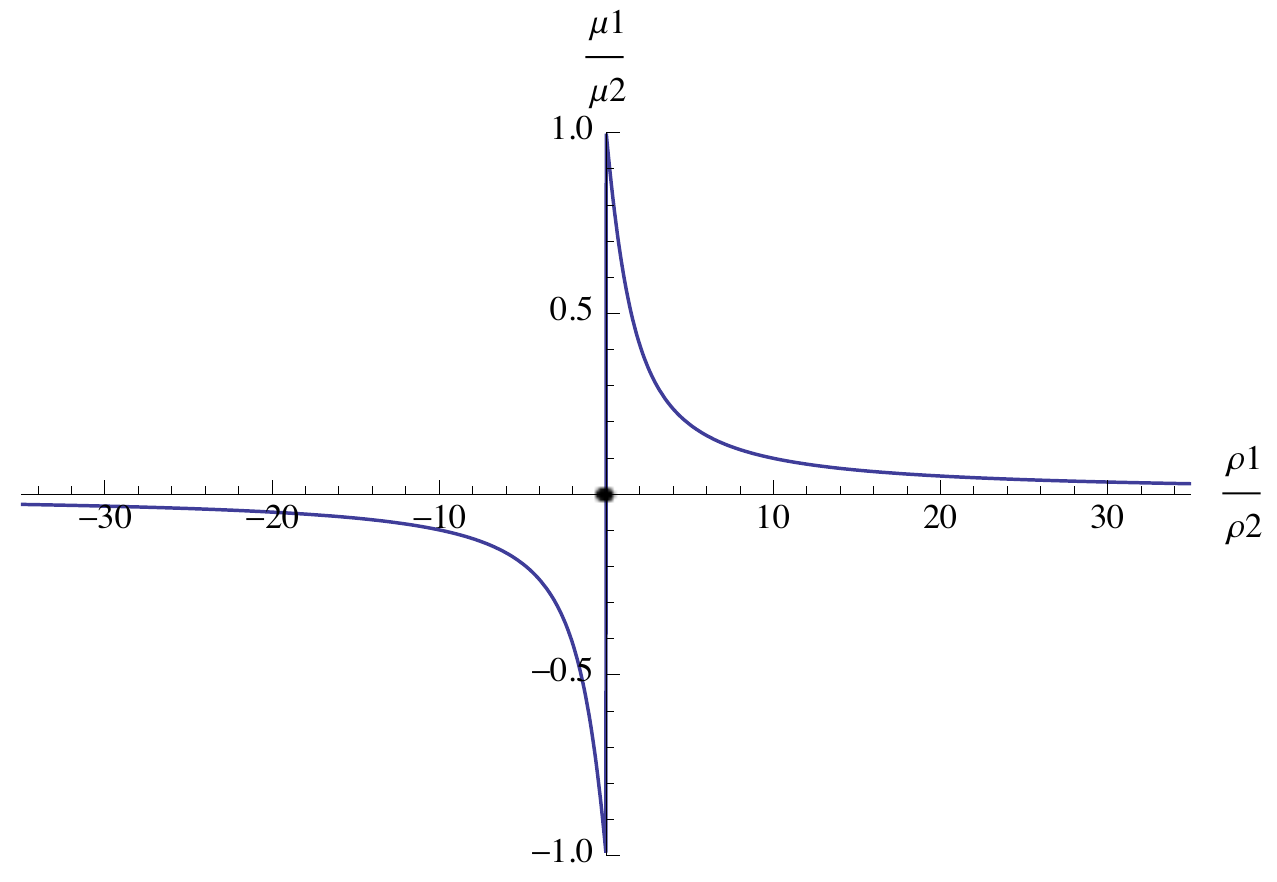}
\caption{The ratio of chemical potentials $\mu_1/\mu_2$ compared with the charge density ratios $\rho_1/\rho_2$. As $\rho_1$ approaches zero from above, $\mu_1$ approaches $\mu_2$ from below, but at $\rho_1 = 0$ strictly we have $\mu_1=0$ (heavy dot at origin).
\label{MuRhoFig}}
\end{center}
\end{figure}
In general the chemical potential $\mu_1$ is realized holographically as (proportional to) the difference in potential between the boundary and the horizon: $\mu_1 \propto \Phi_1(\infty) - \Phi_1(r_H)$.
The form \eno{TwoPlusOneSoln}  for $\Phi_1(r)$ is constructed to have $\Phi_1(r_H) =0$, so in general we simply have $\mu_1 \propto \Phi_1(\infty)$. As $Q_1$ approaches zero, one finds $\Phi_1(\infty) \to Q_2/\sqrt{2}L$ with $\Phi_1(r_H) = 0$; this corresponds to the chemical potential $\mu_1$ approaching the value of $\mu_2$ from below:
\eqn{}{
\lim_{Q_1 \to 0, \; T=0} \; {\mu_1 \over \mu_2} \to 1  \,.
}
At $Q_1 =0$ however, $\Phi_1$ degenerates to the constant \eno{Phi1Shift}, and no longer satisfies
$\Phi_1(r_H) = 0$; see figure~\ref{Phi1Fig}.
Since the difference between the boundary and horizon values is zero, the chemical potential $\mu_1$ vanishes, consistent with the original 2QBH limit. Thus the extremal 2QBH is a state with $\mu_1 = 0$, which nonetheless appears as the limit of a set of extremal geometries approaching $\mu_1= \mu_2$.
We display this behavior in figure~\ref{MuRhoFig}, where we plot $\mu_1/\mu_2$ as a function of the charge density ratio $\rho_1/\rho_2$. Note that as one can show 
\eqn{}{
 {Q_1 \over  Q_2}= {\sqrt{2} \rho_1 \over \rho_2}  \,,
}
one can just as easily think of the $x$-axis as being the black hole charge ratio $Q_1/Q_2$ up to a rescaling.

One can  argue that such a jump in a chemical potential as the charge density is varied is indicative of the presence of a gap in the density of states. As a chemical potential is moved through the gap between bands in a semiconductor, for example, no new states are filled and the density remains constant. Only upon reaching the other side of the gap will the density begin to change again as the nonzero density of states causes new states to be occupied. In studying fermionic fluctuations around the 2QBH, we will find evidence for a gap from another perspective.

We note in addition that the fact that $\mu_1/\mu_2$ is decreasing as $\rho_1/\rho_2$ increases is a signal of a negative susceptibility, and correspondingly of a thermodynamic instability. Such an instability is associated with a perturbative instability in the spectrum of bosonic fluctuations. We do not study such instabilities here, but they were considered originally in \cite{Cvetic:1999rb,Gubser:2000ec,Gubser:2000mm}, and subsequent related work has included \cite{Domokos:2007kt,Nakamura:2009tf,Rozali:2012es,Donos:2013gda}.

\subsection{Fermion spectrum and Dirac equation}

The 5D Dirac equation in the class of backgrounds just reviewed  has been studied for general regular backgrounds in  \cite{Faulkner:2009wj}, and was solved in the 2QBH background with bottom-up fermions in \cite{Gubser:2012yb}. The actual spectrum of fermionic fluctuations in ${\cal N}=4$ SYM was obtained in  \cite{DeWolfe:2012uv} by embedding the black brane solutions \eno{Background} into ${\cal N}=8$ gauged supergravity and then studying the fermionic fluctuations of that theory.
The spin-1/2 fermions of maximal gauged supergravity that do not mix with the gravitino all satisfy a linearized equation of the form
\eqn{DiracEqn}{
\left( i \gamma^\mu \nabla_\mu -  m(\phi) +g q_1 \gamma^\mu a_\mu +gq_2 \gamma^\mu A_\mu + i p_1 e^{-2\phi\over \sqrt{6}} f_{\mu\nu} \gamma^{\mu\nu} + i p_2 e^{\phi \over \sqrt{6}} F_{\mu\nu} \gamma^{\mu\nu} \right) \chi = 0 \,,
}
where the mass term depends on the scalar,
\eqn{MassFunction}{
m(\phi) \equiv g \left( m_1 e^{-{\phi \over \sqrt{6}}} + m_2 e^{2 \phi \over \sqrt{6}} \right) \,,
}
with $m_1$, $m_2$, $q_1$, $q_2$, $p_1$, and $p_2$ rational numbers and $g \equiv 2/L$. The list of so-called ``maximal" mode fermions found in \cite{DeWolfe:2012uv} is
\begin{equation}
\begin{tabular}{|c|c|c|c|c|c|c|c|} \hline
$\chi^{q_a q_b q_c}$ &Dual operator &$m_1$& $m_2$ & $q_1$& $q_2$  & $p_1$ & $p_2$ \\ \hline
$\chi^{({3 \over 2}, {1 \over 2}, {1 \over 2})}$ &$\lambda_1 Z_1$ &$-{1 \over 2}$& ${3 \over 4}$ & ${3 \over 2}$& $1$ & $- {1 \over 4}$& ${1 \over 2}$ \\
$\chi^{({3 \over 2}, -{1 \over 2}, -{1 \over 2})}$ & $\lambda_2 Z_1$&$-{1 \over 2}$& ${3 \over 4}$  & ${3 \over 2}$& $-1$ & $- {1 \over 4} $& $-{1 \over 2}$\\
$\bar\chi^{({3 \over 2}, -{1 \over 2}, {1 \over 2})}$  , $\bar\chi^{({3 \over 2}, {1 \over 2}, -{1 \over 2})}$ & $\overline\lambda_3 Z_1$, $\overline\lambda_4 Z_1$&${1 \over 2}$& $-{3 \over 4}$  & ${3 \over 2}$& $0$ & ${1 \over 4} $&$0$\\ \hline
$\chi^{({1 \over 2}, {3 \over 2}, {1 \over 2})}$, $\chi^{({1 \over 2}, { 1\over 2}, {3 \over 2})}$ & $\lambda_1 Z_2$, $\lambda_1 Z_3$&${1 \over 2}$& $-{1 \over 4}$  & ${1 \over 2}$& $2$ & ${1 \over 4} $& $0$\\
$\bar\chi^{(-{1 \over 2},  {3 \over 2}, {1 \over 2})}$, $\bar\chi^{(-{1 \over 2},  {1 \over 2}, {3 \over 2})}$ & $\overline\lambda_2 Z_2$, $\overline\lambda_2 Z_3$&$-{1 \over 2}$& ${1 \over 4}$ & $-{1 \over 2}$& $2$ & ${1 \over 4}$& $0$ \\ 
$\chi^{(-{1 \over 2},  {3 \over 2}, -{1 \over 2})}$, $\chi^{(-{1 \over 2}, -{1 \over 2},  {3 \over 2})}$& $\lambda_3 Z_2$, $\lambda_4 Z_3$&${1 \over 2}$& $-{1 \over 4}$& $-{1 \over 2}$& $1$ & $- {1 \over 4}$& $- {1 \over 2}$\\
$\bar\chi^{({1 \over 2}, -{1 \over 2},  {3 \over 2})}$, $\bar\chi^{({1 \over 2},  {3 \over 2}, -{1 \over 2})}$& $\overline\lambda_3 Z_3$, $\overline\lambda_4 Z_2$& $-{1 \over 2}$& ${1 \over 4}$ & ${1 \over 2}$& $1$ & $-{1 \over 4}$& ${1 \over 2}$ \\ \hline
\end{tabular}
\end{equation}
Fermions are labeled by the three Cartan charges of $SO(6)$, with $\chi$ and $\bar\chi$ dual to left- and right-handed spinors in four dimensions. The second column describes the composition in terms of the ${\cal N}=4$ gaugini $\lambda_a$, $a=1 \ldots 4$, and adjoint scalars $Z_j \equiv X_{2j - 1} + i X_{2j}$, $j = 1,2,3$; the trace over the gauge group is implied.
Antiparticles with opposite sign charges exist for each row in the table as well. 

We note that, with the exception of the fermion neutral under $q_2$, the fermions come in pairs, the two having the same charge under one of the gauge fields, and the opposite charges under the other. It is thus straightforward to imagine how a Fermi surface with $\rho_1 = 0$ but $\rho_2 \neq 0$ as in the 2QBH could exist: one can have equal numbers of excitations with opposite charges under $a_\mu$ but the same charge under $A_\mu$ filling the ground state.

A few relations can be seen to hold for each fermion: $q_1$ is half-integer quantized, the magnitude of $p_1$ is fixed,
\eqn{p1Relation}{
|p_1| = {1 \over 4} \,,
}
and the parameters $m_2$, $q_1$ and $p_1$ are related,
\eqn{ChargeRelation}{
m_2 = - 2 q_1 p_1 \,.
}
In section~\ref{6DSec}, we will demonstrate that all these relations are required by the consistency of lifting the fermion to six dimensions, and show that the $q_1$ charge corresponds to Kaluza-Klein momentum in the sixth dimension.

One may process the Dirac equation as follows. First we rescale the spinor $\chi$ as
\eqn{ChiToPsi}{
\chi = e^{-2A} h^{-1/4} e^{-i\omega t + i k x} \Psi \,,
}
where $\omega$ is the frequency and $k$ is the spatial momentum, chosen to lie in the $x$-direction,
and the $e^{-2A} h^{-1/4}$ factor exactly cancels the spin connection term coming from $\nabla_\mu$ in \eno{DiracEqn}.
Next, each field $\Psi$ is a four-component spinor; by suitably choosing the Clifford basis, one may decompose this into a pair of two-component spinors $\psi_\alpha$, $\alpha = 1,2$ which decouple from one another  \cite{Faulkner:2009wj, DeWolfe:2012uv}. The Dirac equation for each two-component spinor is then of the form
\eqn{TwoCompDirac}{
\left( \partial_r + X \sigma_3 + Y i \sigma_2 + Z \sigma_1 \right) \psi_\alpha = 0\,,
}
where
\eqn{XYZDefs}{
X \equiv  {m e^B \over \sqrt{h}} \,, \quad \quad Y \equiv- {e^{B-A} \over \sqrt{h}} u \,, \quad\quad
Z \equiv -{e^{B-A} \over \sqrt{h}}((-1)^\alpha k -v) \,,
}
and we have defined (following \cite{Faulkner:2009wj, DeWolfe:2012uv})
\eqn{uandv}{
u \equiv {1 \over \sqrt{h}} \left(\omega + g q_1 \Phi_1 + g q_2 \Phi_2 \right) \,, \quad \quad
v \equiv 2 e^{-B} \left( p_1 e^{-{2 \phi \over \sqrt{6}}} \partial_r \Phi_1 + p_2 e^{\phi \over \sqrt{6}} \partial_r \Phi_2 \right)\,.
}
Let us write each two-component spinor as
\eqn{}{
\psi_\alpha = \begin{pmatrix}\psi_{\alpha-} \cr\psi_{\alpha+}\end{pmatrix} \,.
} 
It what follows we will suppress the label $\alpha$ on $\psi$; it is clear from \eno{XYZDefs} that the solution for one two-component spinor will be identical to that for the other with $k \to -k$.

We will now pass to decoupled second-order equations, in a fashion outlined in \cite{Gubser:2012yb}.\footnote{For the numerical solutions of section~\ref{Nequals4Sec} we used a different, equivalent second-order equation, derived in \cite{DeWolfe:2012uv} and discussed in  appendix~A. The choice given here is more convenient for the analysis of sections~\ref{RegularSec} and \ref{2QBHSec}.} Define the combinations
\eqn{}{
U_{ \pm} \equiv \psi_{ -} \pm i \psi_{ +} \,.
}
The first order equations then take the form
\eqn{}{
U_{ -}' + i Y U_{ -} &= (-X+iZ) U_{ +} \,,  \cr
U_{ +}' - i Y U_{ +} &= (-X-iZ) U_{ -} \,, 
}
and lead to the second-order equations
\eqn{DiracSecondOrder}{
U_-'' + p \, U_-' + (i Y'-X^2 + Y^2 - Z^2 + i  Y p) U_- &= 0 \,, \cr
U_+'' + \bar{p} \, U_+' + (-i Y'-X^2 + Y^2 - Z^2 - i  Y \bar{p}) U_+ &= 0 \,,
}
where
\eqn{pDef}{
p \equiv - \partial_r \log(-X + i Z) \,. 
}
To calculate a retarded Green's function, one solves this equation with infalling boundary conditions imposed at the horizon.
The response is then read off from the behavior of solutions near the boundary $r \to \infty$,
\eqn{BoundaryScalings}{
\psi_+ \sim A(\omega, k) r^{mL} + B(\omega, k) r^{-mL - 1} \,, \quad \quad
\psi_- \sim C(\omega, k) r^{mL-1} + D(\omega, k) r^{-mL} \,,
}
with $mL \equiv 2(m_1 + m_2)$. For $m > 0$, $A$ is the source term, and $D$ the response (the case $m<0$  exchanges their roles, corresponding to a dual fermion of opposite chirality). The Green's function is then defined as the ratio of the response to the source,
\eqn{FullGreen}{
G_R = {D \over A}  \,.
}
A Fermi surface is  identified as the momentum $k_F$ where there is a pole in the retarded Green's function at zero $\omega$,
\eqn{FermiSurface}{
A(\omega = 0, k= k_F) \equiv 0 \,.
}
To make progress, one should study the near-horizon behavior of the Dirac equation, where the infalling condition is imposed. We first review the well-known ``regular" cases in the next section, before turning to the 2QBH case in the section following.

\section{Green's functions for regular extremal black holes}
\label{RegularSec}

In this section we review the near-horizon analysis of the Dirac equation in ``regular" extremal black holes, the calculation of the IR Green's function and using it to obtain an expression for the full Green's function, and the extraction of the dispersion relation for fluctuations. This will be useful as a warm-up to the case of the two-charge black hole. This analysis was originally performed from a bottom-up perspective by \cite{Faulkner:2009wj}, and was extended to the maximal gauged supergravity Dirac equation \eno{DiracEqn} in \cite{DeWolfe:2012uv}.

Generic (2+1)-charge extremal black holes are ``regular", having a double pole at the horizon but no singularity there.
One has in the near-horizon limit $r \to r_H$,
\eqn{}{
e^A \to k_0 \,, \quad \quad e^B \to {\tau_0 L_2 \over k_0} \,, \quad \quad h \to \left( \tau_0 \over k_0\right)^2 (r - r_H)^2 \,, \quad \quad
\Phi_i \to \beta_i (r -r_H) \,, \quad \quad \phi \to \phi_0 \,,
}
where $k_0$, $\tau_0$, $L_2$, $\beta_i$ and $\phi_0$ are all constants defined in \cite{DeWolfe:2012uv}, leading to the near-horizon geometry $AdS_2 \times \mathbb{R}^3$ \cite{Faulkner:2009wj},
\eqn{}{
ds^2 =  -\tau_0^2 (r - r_H)^2 dt^2  + { L_2^2 dr^2 \over (r - r_H)^2} + k_0^2 d\vec{x}^2\,.   
}
We see in the metric everything is regular and approaches a constant except for the horizon function $h$; the gauge fields have a single zero.
The second-order Dirac equation \eno{DiracSecondOrder} then has the form as $r \to r_H$:
\eqn{RegularSecondOrder}{
 U'' + \left({1 \over r - r_H} + \ldots \right)  U' + \left(  {L^4 (Q_2^4 - r_H^4) \omega^2 \over 16 (2 Q_2^2 - r_H^2)(r - r_H)^4} + {\# \omega \over (r - r_H)^3} 
- {\nu^2 \over (r-r_H)^2} + \ldots \right) U= 0\,,
}
where we have neglected terms of order $\omega^2/(r-r_H)^3$ and $\omega/(r-r_H)^2$ as well as $(r-r_H)^{-1}$  in the no-derivative term, $\#$ is a constant whose form we will not record, and $\nu^2$ is given as
\eqn{TwoPlusOneNuk}{
\nu^2 = { (m_1 (1 - \mu_R)^2 + m_2 \mu_R^2)^2 \over (1 - \mu_R^4)} + {\tilde{k}^2 \over \mu_2^2}{1  \over 2(1 + \mu_R^2)} - {(\sqrt{2} q_1 \mu_R^3 + q_2 (1 - \mu_R^2) )^2 \over 4 (1 - \mu_R^2)(1 + \mu_R^2)^2}\,,
}
with $\mu_R \equiv \mu_1/\mu_2$ and 
where\footnote{Note there is an overall sign typo in the second line of (123) in \cite{DeWolfe:2012uv}.}
\eqn{}{
\tilde{k} \equiv k -(-1)^\alpha (2 p_1 \mu_1 + \sqrt{2} p_2 \mu_2) \,.
}
One sees that near the horizon, the terms involving $\omega$ are dominant. This is a sign that interesting physics happens near $\omega =0$. We may study the small $\omega$ limit, but for any fixed, tiny $\omega$ there will always be values of $r$ sufficiently close to the horizon that the first two terms in parentheses dominate. Thus the small-frequency and near-horizon limits do not commute. Yet, we must understand the behavior of a mode near the horizon to impose infalling boundary conditions. 
For this reason, we must consider inner and outer regions.

Consider the inner region first. This is the region designed to take into account the values of the radial coordinate that are so close to the horizon that $\omega$ cannot be ignored, even though it is small.
For the inner region, we scale $\omega$ and $r - r_H$ at the same time: $r \to r_H$, $\omega \to 0$ with $\omega/(r - r_H)$ held fixed.  The inner region equation then takes precisely the form of \eno{RegularSecondOrder} with all neglected terms scaled to zero:
\eqn{InnerRegionEqn}{
 U'' + {1 \over r - r_H}  U' + \left(   {L^4 (Q_2^4 - r_H^4) \omega^2 \over 16 (2 Q_2^2 - r_H^2)(r - r_H)^4} + {\# \omega \over (r - r_H)^3} 
- {\nu^2 \over (r-r_H)^2}  \right) U= 0\,.
}
In the inner region we then solve this equation for all $r$ with fixed $\omega$.
In general there is a solution in the inner region that is purely infalling as $r \to r_H$; this can be obtained explicitly in terms of Whittaker functions, but we will not record the form here.

The near-boundary limit of $AdS_2$ is where the inner region glues onto the outer region, corresponding to $r - r_H$ large (with $\omega$ kept fixed), which becomes 
\eqn{RegularInnerNearBdy}{
U'' + {1 \over r - r_H}  U' - {\nu^2 \over (r-r_H)^2}   U= 0\,.
}
This equation has power law solutions, 
\eqn{PowerLaws}{
U \sim (r - r_H)^{- {1 \over 2} \pm \nu}\,.
}
In general the solution of the whole inner region that is infalling at small $r-r_H$ can be expanded at large $r - r_H$ in some linear combination of \eno{PowerLaws},
\eqn{AdS2Bdy}{
U \sim (r - r_H)^{- {1 \over 2} + \nu} + {\cal G}(\omega) (r - r_H)^{- {1 \over 2} - \nu} \,,
}
where the relative weighting ${\cal G}(\omega)$ between the two solutions depends on $k$ and the other parameters as well; thinking of the two terms in \eno{AdS2Bdy} as the source and response in an $AdS_2$ fluctuation, we may consider ${\cal G}(\omega)$ to be an $AdS_2$ Green's function. Since the $AdS_2$ region corresponds to the IR part of the full geometry, this is also called the IR Green's function. Note that the only dependence on $\omega$ in \eno{AdS2Bdy} is in ${\cal G}(\omega)$.
The full form of ${\cal G}(\omega)$ is recorded in \cite{Faulkner:2009wj};
for small $\omega$ it takes the form,
\eqn{RegularCalG}{
{\cal G}(\omega) = |c(k)| e^{i \gamma_k} (2\omega)^{2 \nu} \,,
}
for real quantities $|c(k)|$ and $\gamma_k$.

Now consider the outer region. Here by definition $r - r_H$ is large enough that the $\omega$ terms in \eno{RegularSecondOrder} can be neglected. The near-horizon equation for the outer region is then
\eqn{RegularOuterNearHor}{
U'' + {1 \over r - r_H}  U' - {\nu^2 \over (r-r_H)^2}   U= 0\,,
}
which precisely matches \eno{RegularInnerNearBdy}; the two regions have the same solutions on the overlap region. Thus in the outer region we have in general $\omega \to 0$ solutions $\eta^0_\pm$ with near-boundary behavior
\eqn{NearBoundary}{
\eta^0_\pm \to (r - r_H)^{- {1 \over 2} \pm \nu} \,,
}
and thus the solution that has infalling boundary conditions in the outer region takes the form
\eqn{}{
U \sim \eta_+^0 + {\cal G}(\omega) \eta_-^0 \,.
}
In general as we move away from $\omega \to 0$ we will have corrections
\eqn{}{
\eta_\pm = \eta^0_\pm + \omega \eta^1_\pm + \ldots\,,
}
and thus the small-$\omega$ solutions are
\eqn{RegularOuterSolns}{
U \sim \eta_+ + {\cal G}(\omega) \eta_- \,.
}
Both components of the two-component spinors take this generic form. Thus passing from $U_\pm$ to $\psi_\pm$ and expanding near the boundary, we obtain a general expression for the retarded Green's function near $\omega = 0$ as
\eqn{FullGreenAgain}{
G = {D \over A} = {b_+^0 + \omega  b_+^1 + \ldots + {\cal G}(\omega) (b_-^0 + \omega b_-^1 + \ldots)
\over a_+^0 + \omega a_+^1 + \ldots + {\cal G}(\omega) (a_-^0 + \omega a_-^1 + \ldots)} \,,
}
where the $a_\pm^i$ and $b_\pm^i$ are $k$-dependent.
Calculating the coefficients $a_\pm^i$ and $b_\pm^i$ in general requires numerical solution of the Dirac equation over the entire space, but the IR Green's function ${\cal G}(\omega)$ and the associated parameter $\nu$ may be determined solely from the IR physics. 

The quantity $\nu^2$ may be positive or negative depending on the parameters, including $k$, and thus $\nu$ may be real or imaginary. When $\nu$ is imaginary, the boundary conditions \eno{NearBoundary} become complex, and it is not possible to find a Fermi surface as one cannot make the denominator  of the Green's function \eno{FullGreenAgain} vanish at $\omega =0$ while keeping $k$ real; in general Im $G_R(\omega = 0, k) \neq0$. In this regime, called the oscillatory region, the Green's function is periodic in $\log \omega$, and one has gapless excitations for a range of $k$  \cite{Faulkner:2009wj}. This behavior is associated to an instability in the IR of the geometry towards pair production due to the strong electric field \cite{Pioline:2005pf}, and the corresponding behavior in scalars leads to unstable modes.

When $\nu$ is real, on the other hand, we may find a Fermi surface at a particular Fermi momentum $k = k_F$; 
determining $k_F$ requires solving the Dirac equation out from the horizon to the boundary to find \eno{FermiSurface}, and thus the value of the Fermi momentum requires knowledge of the UV physics.
One then has $a_+^0(k_F) =0$, and near the Fermi surface the Green's function takes the form
\eqn{GreensFunction}{
G_R(k, \omega) \sim {h_1 \over k_\perp - {1 \over v_F} \omega + \ldots - h_2 e^{i \gamma_{k_F}} (2 \omega)^{2 \nu_{k_F}}} \,.
}
Here $k_\perp \equiv k - k_F$, and we have plugged in the form \eno{RegularCalG} for ${\cal G}(\omega)$. The real constants $h_1$, $h_2$ and $v_F$ are more difficult to compute, but once $k_F$ is known some physics can be extracted just from knowledge of $\nu_{k_F}$ and $\gamma_{k_F}$. 

The dispersion relation for excitations near the Fermi surface is defined by the values of $(k_\perp, \omega)$ for which the denominator of \eno{GreensFunction} vanishes, and thus has the form
\eqn{}{
k_\perp = {1 \over v_F} \omega + \ldots + h_2 \cos \gamma_{k_F} (2 \omega)^{2 \nu_{k_F}}
+ i h_2 \sin \gamma_{k_F} (2 \omega)^{2 \nu_{k_F}} \,.
}
Since $v_F$ is real, 
the properties of the IR Green's function ${\cal G}(\omega)$ determine the leading imaginary part in the dispersion relation to scale as $\omega^{2 \nu_{k_F}}$; if we have $\nu_{k_F} < 1/2$ this is the leading real part as well, beating the $\omega/v_F$ term. For $\nu_{k_F} > 1/2$ we have a Fermi liquid; for $\nu_{k_F} < 1/2$, a non-Fermi liquid, with marginal Fermi liquid sitting in between.

For a Fermi liquid, fluctuations near the Fermi surface become asymptotically stable as the fluctuation energy goes to zero.
For a non-Fermi liquid, on the other hand, the ratio of the width $\Gamma$ to the energy $\omega_*$ of a fluctuation near the Fermi surface approaches the constant
\eqn{}{
{\Gamma \over \omega_*} = \tan {\gamma_{\nu_F} \over 2 \nu_F} \,.
}
The residue of the (would-be) quasiparticle pole is then a constant for a Fermi liquid, and for the non-Fermi liquid scales as
\eqn{}{
Z \sim \left( k_\perp \right)^{{1 \over 2 \nu_{k_F} } - 1} \,.
}
In  \cite{DeWolfe:2012uv}, the fermionic fluctuations of all spin-1/2 modes not mixing with the gravitino were studied for all values of $\mu_R$ in the extremal (2+1)QBH backgrounds. A number of Fermi surface singularities were discovered, all associated to non-Fermi liquids.

\section{Green's functions for the extremal two-charge black hole}
\label{2QBHSec}

We now turn to our main interest, the extremal two-charge black hole (2QBH). The horizon in this case is at $r =0$, but this case is different because it is singular as $r \to 0$ as well. The functions $X$, $Y$ and $Z$ that appear in the Dirac equation, expanded near $r \to 0$, are
\eqn{}{
X = {a \over r^2} + b \,, \quad \quad
Y = {c \omega \over r^2} + d \omega + f \,, \quad \quad
Z = {P \over r} \,, 
}
where $a$, $b$, $c$, $d$, and  $f$ are constants depending on the background and the type of fermion,
\eqn{Constants}{
a = m_2 \mu L^2 \,, \quad  b = {1 \over \mu L^2} \left( 2 m_1 + {3 \over 2} m_2 \right) \,, \quad
c = - {L^2 \over 2} \,, \quad d = - {1 \over 2 \mu^2 L^2} \,, \quad 
 f = - {q_2 \over \sqrt{2}} {1 \over \mu L^2} \,, 
}
and $P$ includes the spatial momentum,
\eqn{}{
P = - (-1)^\alpha {k \over \mu} + \sqrt{2} p_2 \,.
}
We are neglecting terms of ${\cal O}(r^2)$ from $X$ and $Y$, and ${\cal O}(r)$ from $Z$. One can show that the quantity $p$ defined in \eno{pDef} is
\eqn{}{
p = {2 \over r} + {i P \over a} + {\cal O}(r) \,,
}
which implies that
\eqn{}{
iY' + i Y p  = -{c \omega P \over a r^2} + {\cal O}\left(1 \over r\right)\,,
}
where two terms of ${\cal O}(r^{-3})$ canceled. 
We then find that the near-horizon second-order equation \eno{DiracSecondOrder} takes the same form for either $U_\pm$, with differences subleading in $r$, and thus can also be written as an equation
for either $\psi_\pm$, as
\eqn{TwoQEqn}{
\psi'' + \left( {2 \over r} + \ldots \right) \psi' + \left( {c^2 (\omega^2 - \Delta^2) \over r^4} + {{1 \over 4} - 4\nu^2  + {\cal O}(|\omega| - \Delta) \over r^2} + \ldots  \right) \psi =0 \,,
}
where we have defined the scale $\Delta$, closely related to the chemical potential,
\eqn{}{
\Delta \equiv \left| a \over c \right| = 2 |m_2| \mu \,,
}
and the quantity $\nu^2$ is
\eqn{NuSquared}{
\nu^2 \equiv {1 \over 4} \left(P - {1 \over 2} \sgn(\omega m_2)\right)^2
+ {1 \over 4} m_2^2 +  m_1 m_2 - {\sgn(\omega m_2)\over 2 \sqrt{2}} q_2 m_2\,.
}
This quantity has been given the same name as \eno{TwoPlusOneNuk} for the (2+1)QBH case; as we shall see, they will play an analogous role, and even coincide in the appropriate limit.

The equation \eno{TwoQEqn} has obvious similarities to the (2+1)QBH case \eno{RegularSecondOrder}, but differences as well. Most salient is that both share $1/(r-r_H)^4$ terms that dominate near the horizon, but are suppressed by factors of energy. For the (2+1)QBH case, it was near $\omega = 0$ that the $1/(r-r_H)^4$ terms were suppressed, leading to the need for inner and outer regions there. For the extremal 2QBH, this happens instead at $|\omega| = \Delta$. This difference lies at the heart of the novel structure for this case.

\subsection{Solutions away from $|\omega| = \Delta$}

As long as $\omega^2 - \Delta^2$ is not small, the first term in parentheses in \eno{TwoQEqn} dominates over all other zero-derivative terms at small $r$, and we get
\eqn{TwoQEqnAway}{
 \psi'' + {2\over r} \psi' + {c^2 (\omega^2 - \Delta^2) \over r^4} \psi =0 \,.
}
There is no need in these cases to worry about inner and outer regions.
The solutions of \eno{TwoQEqnAway} were considered in \cite{DeWolfe:2012uv}. Let us define the quantity $\epsilon$ which measures the deviation of $|\omega|$ away from $\Delta$,
\eqn{}{
\epsilon^2 \equiv {\omega^2 \over \Delta^2}- 1  \,.
}
Then for $|\omega| > \Delta$, we have the complex solutions
\eqn{InOutFalling}{
\psi \sim \exp \left( \pm{ i \Delta L^2 \epsilon  \over 2 r}\right) \,,
}
corresponding to infalling and outgoing waves at the horizon. In this case the usual infalling boundary conditions are easy to impose, and since the infalling wave is complex, the corresponding fluctuation is complex as well.

On the other hand, for $|\omega| < \Delta$ we have instead growing and dying solutions,
\eqn{}{
\psi \sim \exp \left( \pm{ \Delta L^2 |\epsilon|  \over 2 r}\right) \,.
}
As was noted in \cite{DeWolfe:2012uv}, there is no obvious solution here corresponding to an infalling wave. Instead, it was suggested there to adopt the standard boundary condition of Euclidean AdS/CFT, keeping the regular solution at the horizon, corresponding to keeping the dying exponential and removing the diverging exponential. Choosing the non-diverging solution is a real boundary condition, and since the differential equation is real, we will in this case have purely {\em real} solutions to the fluctuation equations.
The Fermi surface can be defined as usual \eno{FermiSurface} as the vanishing of the retarded Green's function at $\omega = 0$; however, since this lies in the $|\omega| < \Delta$ region, to do this we must accept this modified prescription for the boundary conditions.  It appears that the Fermi surface is surrounded by an energy gap of size $2\Delta$, within which the modes are purely real; as we shall describe, the corresponding fluctuations have no decay width and so are stable. In the next subsection we look at the region $|\omega| \approx \Delta$, where this  unusual behavior crosses back over to more typical behavior. 

\subsection{Solutions near $|\omega| = \Delta$}

Looking at the near-horizon equation \eno{TwoQEqn}, it is evident that the kind of ``interesting" behavior that showed up in the regular case  \eno{RegularSecondOrder} at $\omega = 0$ happens here instead at
$|\omega|  \to  \Delta$; this is the case for which the first term in parentheses is suppressed except for very close to the horizon.
The analog of studying small $\omega$ for the regular cases is to study small $\epsilon$, that is, to study energies near $\Delta$. Recall that $\omega =0$ already corresponds to fluctuations at the Fermi surface $\varepsilon_F = q_2 \mu$; so we are now discussing fluctuations around a scale $\Delta = 2 |m_2| \mu$ above or below the Fermi energy.

Let us analyze \eno{TwoQEqn} for energies near $\Delta$. As with regular cases near $\omega = 0$, 
we have the issue that for fixed small $\epsilon$ there will always be sufficiently small $r$ where the $1/r^2$ term dominates, and again we can solve this with inner and outer regions. For the outer region, $r$ is large enough that the relevant term can be neglected, and we set $\omega = \Delta$ --- that is, $\epsilon =0$. The second-order equation in the outer region then has the near-horizon limit,
\eqn{TwoQOuterFirst}{
 \psi'' +{2 \over r} \psi' +   {{1 \over 4} - 4 \nu^2  \over r^2} \psi =0 \,.
}
The solutions to \eno{TwoQOuterFirst} are power laws,
\eqn{InnerBoundaryPower}{
\psi \sim r^{- {1 \over 2} \pm 2 \nu} \,.
}
Meanwhile the inner region is defined by taking $r\to0$, $\epsilon \to 0$, with $\epsilon/r$ fixed. We obtain the inner region equation,
\eqn{TwoQInner}{
 \psi'' + {2 \over r} \psi' + \left( {a^2\epsilon^2 \over r^4} + {{1 \over 4} - 4 \nu^2 \over r^2}   \right) \psi =0 \,.
}
The near-boundary limit of the inner region involves $r \to \infty$ with $\epsilon$ fixed, and results in the same equation \eno{TwoQOuterFirst} as the near-horizon limit of the outer region; as expected there is a match in the transition region.

We can solve the inner region equation in general, obtaining  
$r^{-1/2}$ times Bessel functions. For  $|\omega| > \Delta$ we have
\eqn{Bessels}{
\psi \sim{1 \over \sqrt{r}}  \, J_{2\nu}\!\left( \Delta L^2 \epsilon  \over 2 r\right) , \; {1 \over \sqrt{r}}  \, Y_{2\nu}\!\left( \Delta L^2 \epsilon  \over 2 r\right) \,, \quad \quad \omega > \Delta \,.
}
while for $\omega < \Delta$, we have imaginary $\epsilon$, and the above become modified Bessel functions,
\eqn{ModifiedBessels}{
\psi \sim {1 \over \sqrt{r}} \, K_{2\nu}\!\left( \Delta L^2 |\epsilon|  \over 2 r\right) , \;  {1 \over \sqrt{r}}  \, I_{2\nu}\!\left( \Delta L^2 |\epsilon|  \over 2 r\right) \,, \quad \quad \omega < \Delta \,.
}
For $\omega > \Delta$, one may impose infalling boundary conditions at the horizon by choosing the combination of \eno{Bessels} to be the Hankel function of the first kind,
\eqn{Hankel}{
\psi = {1 \over \sqrt{r}}  H^{(1)}_{2\nu}\!\left(\Delta L^2 \epsilon  \over 2 r  \right)  \to \sqrt{2 \over \epsilon \pi} e^{- {i \pi \over 4} - {2 \pi i \nu \over 2}} \exp \left( i \Delta L^2 \epsilon  \over 2r \right) \,, 
}
which indeed takes in the infalling form at $r \to 0$, matching \eno{InOutFalling}. We can take this solution and 
 examine the inner region near-boundary ($r \to \infty$) behavior and extract the ``IR Green's function" ${\cal G}$.\footnote{In the regular case, the term IR Green's function was well-motivated by the fact that the near-horizon region was $AdS_2$ and hence had its own holographic interpretation. We will continue to use the terminology; later we will discuss the relation of this near-horizon region and $AdS_3$.}
 Consider for simplicity the case of non-integral $2\nu$; this will be generic except for  special values of the momentum. As $x \to 0$ we have,
\eqn{}{
H_{2\nu}^{(1)}(x) = - {i \Gamma(2\nu) \over \pi} \left( x \over 2 \right)^{-2\nu} + \ldots  - {i \Gamma(-2\nu) \over \pi} e^{-2 \pi i \nu} \left( x \over 2 \right)^{2\nu} + \ldots \,,
}
where each leading term is corrected by a power series in even powers of $x$; since $2\nu$ is non-integral here these do not overlap.  Thus we see that
\eqn{}{
\psi = - {i \over \pi} \left( 
 \Gamma(2\nu) \left(\Delta L^2 \epsilon \over 4 \right)^{-2\nu} r^{{2\nu} - {1 \over2}} 
+ \ldots + \Gamma(-2\nu) e^{-2 \pi i\nu} \left( \Delta L^2 \epsilon \over 4 \right)^{2\nu} r^{-2\nu - {1 \over2}} 
\right) \,.
}
These power laws indeed match the two solutions for the near-boundary limit of the inner region from \eno{InnerBoundaryPower}. The ratio between the $r^{-2\nu - {1 \over 2}}$ and $r^{2\nu - {1 \over 2}}$ terms is
\eqn{GPlus}{
{\cal G}_+(\epsilon) &= e^{-2 \pi i\nu} {\Gamma(-2\nu) \over \Gamma(2\nu) } \left( \Delta L^2 \epsilon \over4 \right)^{4\nu} \cr
&= e^{-2 \pi i \nu} {\Gamma(-2\nu) \over \Gamma(2\nu) } \left( L^4 (\omega^2 - \Delta^2) \over16 \right)^{2 \nu} \,,
} 
where we use the label $+$ to indicate $\omega > \Delta$.  For the case of $\nu$ real, we take $\nu>0$ and then the $r^{2\nu - {1 \over 2}}$ term always dominates over $r^{-2\nu - {1 \over 2}}$, so we may think of them as the source
 and the response  respectively; then ${\cal G}(\epsilon)$ may usefully be thought of as the IR Green's function.

For $|\omega|-\Delta$ small, we have
\eqn{}{
{\cal G}_+(\epsilon) \approx e^{-2 \pi i \nu} {\Gamma(-2\nu) \over \Gamma(2\nu) } \left( L^4 \Delta (|\omega| - \Delta) \over8 \right)^{2 \nu} \,.
}
Comparing this to the form \eno{RegularCalG} for the regular cases, we see both scale like the deviation from the ``special energy" to the $2 \nu$ power. Furthermore, in this case for real $\nu$ the phase of ${\cal G}$ is simply
\eqn{GammaPhase}{
\gamma \equiv \arg {\cal G}= -2 \pi \nu \,. 
}
Consider now the case of $\omega < \Delta$, so $\epsilon$ is imaginary; here we will give results in terms of $|\epsilon| = \sqrt{1 - \omega^2/\Delta^2}$.
Now the solutions \eno{ModifiedBessels} are purely real, and consequently they do 
not show infalling/outgoing behavior at the horizon; instead, one solution is regular there and one is divergent. Not having the infalling prescription available, we will instead follow the prescription used for Euclidean AdS/CFT and choose the regular solution, which is
\eqn{BelowTwo}{
\psi = {1 \over \sqrt{r}} \,K_{2\nu} \!\left(\Delta L^2 |\epsilon|  \over 2 r   \right) \,.
}
The expansion near the boundary  is
\eqn{}{
\psi =  {\Gamma(2\nu) \over 2} \left( \Delta L^2 |\epsilon| \over 4 \right)^{-2\nu} r^{2\nu - {1 \over 2}} + \ldots  + {\Gamma(-2\nu) \over 2} \left(\Delta L^2 |\epsilon| \over 4 \right)^{2\nu} r^{-2\nu - {1 \over 2}} + \ldots  \,,
}
and so now the IR Green's function is 
\eqn{GMinus}{
{\cal G}_-(\epsilon) = {\Gamma(-2\nu) \over \Gamma(2\nu)} \left( \Delta L^2 |\epsilon|  \over 4 \right)^{4 \nu} 
= {\Gamma(-2\nu) \over \Gamma(2\nu)} \left( L^4( \Delta^2 - \omega^2)  \over 16 \right)^{2 \nu}\,,
}
which for $\Delta - |\omega|$ small is
\eqn{GMinusSmall}{
{\cal G}_-(\epsilon) \approx {\Gamma(-2\nu) \over \Gamma(2\nu)} \left( L^4 \Delta ( \Delta - |\omega|)  \over 8 \right)^{2 \nu}\,.
}
We are now in a position to further justify the regular prescription: it is the continuation of the infalling prescription as $|\omega|$ moves past $\Delta$, corresponding to  imaginary $\epsilon$. This can be seen using the Bessel function identity:
\eqn{BesselIdentity}{
K_{2\nu}(x) = {\pi \over 2} i^{2\nu + 1} H_{2\nu}^{(1)}\!(ix)\,,
}
showing the expressions for ${\cal G}_+(\epsilon)$ and ${\cal G}_-(\epsilon)$ are the same formula, with $\epsilon$ continued from real to imaginary values; we drop the distinction from here on. This provides further confidence in our choice of boundary condition for the $|\omega| < \Delta$ region.

As in the regular cases, the outer region solution will match to the inner solution on their overlap, and thus the continuation of our boundary condition to the outer region gives solutions 
\eqn{}{
\psi = \eta_+ + {\cal G}(\epsilon) \eta_- \,,
}
where $\eta_\pm \sim r^{- {1 \over 2} \pm 2\nu}$ and as in the regular case, the solutions can be extended away from $\epsilon =0$ in a power series in $\epsilon^2$; for small deviations $\epsilon^2$ is proportional to the linear deviation,
\eqn{}{
\epsilon^2 \approx {2 (|\omega| - \Delta) \over \Delta} \,.
}
Meanwhile the outer region near-boundary behavior still has the form \eno{BoundaryScalings}, and since both components have the same near-horizon behavior, we get for the full Green's function near $|\omega| = \Delta$:
\eqn{TwoFullGreen}{
G_R = {D \over A} = {b_+^{(0)} + \epsilon^2  b_+^{(2)} + \ldots + {\cal G}(\epsilon) (b_-^{(0)} + \epsilon^2 b_-^{(2)} + \ldots)
\over a_+^{(0)} + \epsilon^2 a_+^{(2)} + \ldots + {\cal G}(\epsilon) (a_-^{(0)} + \epsilon^2 a_-^{(2)} + \ldots)} \,.
}
Thus the IR Green's function ${\cal G}(\epsilon)$ plays the same role near $|\omega| = \Delta$ as its cousin ${\cal G}(\omega)$ does near $\omega = 0$ in the regular case.

 As with the regular case, there are different situations depending on the sign of $\nu^2$ \eno{NuSquared}. Note that the momentum dependence of $\nu^2$ enters in the positive-definite term $(P \pm 1/2)^2$; this is played against a constant term depending on the fermion in question, which may be negative. If $\nu^2 < 0$, the power law solutions \eno{InnerBoundaryPower} are not real, and we again find an oscillatory region where there are no Fermi surface singularities. For $\nu^2 > 0$, Fermi surfaces may be found.

Note that the form \eno{TwoFullGreen} is not useful near $\omega = 0$, so it does not tell us anything about fluctuations near the Fermi surface. Instead, it tells us about dynamics near the gap $\Delta$. The denominator may vanish for a particular value of $k \equiv k_\Delta$, corresponding to a pole in the Green's function at $\omega = \Delta$. (One may also define a $k_{-\Delta}$ where the denominator vanishes for $\omega = - \Delta$, and $k_\Delta$ and $k_{-\Delta}$ need not coincide.) The spectrum of nearby fluctuations may then be worked out. 

For $|\omega| > \Delta$, the case of complex fluctuations, the behavior of the Greens function near $k = k_{\Delta}$ is
\eqn{}{
G_R \sim {h_1 \over (k- k_\Delta) - {|\omega| - \Delta \over v_F} + \ldots - h_2 e^{- 2 \pi i \nu_{k_{\Delta}}} (|\omega| - \Delta)^{2 \nu_{k_{\Delta}}}} \,,
}
and we find a story very similar to the regular case \eno{GreensFunction}. The energy dependence scales with the quantity $\nu$ in an identical fashion, justifying our choice for the analogous notation. The phase is determined by $\nu$ as well, and once again we will have a Fermi liquid if $\nu_{k_\Delta} > 1/2$, a non-Fermi liquid for $\nu_{k_\Delta} < 1/2$, and a marginal Fermi liquid for $\nu_{k_\Delta} = 1/2$. For the non-Fermi case, the ratio of the width to the energy is
\eqn{PosOmegaDisp}{
{\Gamma \over \omega_*}  = \tan {- 2 \pi \nu_\Delta  \over 2 \nu_\Delta} = 0 \,.
}
Thus unlike the generic regular case, we find that even for cases with $\nu < 1/2$, the dispersion of the modes vanishes as the special energy is approached. We can still consider these non-Fermi liquids, as the real and imaginary parts of the Green's function denominator scale with the same power law, and other quantities such as the residue consequently keep the non-Fermi liquid form
\eqn{}{
Z \sim \left( k- k_\Delta \right)^{{1 \over 2 {\nu_\Delta} } - 1} \,.
}
One might say that $\Gamma/\omega_*$ approaches a constant as with other non-Fermi liquids, but the constant in this case is zero.

For $\omega < \Delta$, we have instead the form for the Green's function,
\eqn{}{
G_R \sim {h_1 \over (k- k_\Delta) + {|\omega| - \Delta \over v_F} + \ldots - h_2 (\Delta - |\omega|)^{2 \nu_{k_{\Delta}}}} \,,
}
which is just the continuation of the first form to negative $|\omega|  - \Delta$.  The primary difference is that everything is real. As a result, the energy $\omega$ of the fluctuation has no imaginary part; the correpsonding modes are stable everywhere for $|\omega| < \Delta$. This connects smoothly to the result \eno{PosOmegaDisp} for $|\omega| > \Delta$.

\subsection{Connection with extremal (2+1)-charge black holes}

In this subsection, we show how the boundary of the stable region $|\omega| \approx \Delta$ can  be probed by studying the limit of Fermi surfaces at $\omega \approx 0$ in extremal (2+1)-charge black holes. When we study results for specific fermions of maximal gauged supergravity in 2QBH backgrounds in the next section, this will connect those results to the results of \cite{DeWolfe:2012uv}.

 As described in section~\ref{GravitySec}, approaching the extremal 2QBH via extremal (2+1)QBHs shifts $\Phi_1$ by a constant \eno{Phi1Shift}. This constant shift, while not a true shift in the chemical potential, does correspond to a simple shift in the zero point of the energy for $q_1$-charged particles. Consider the Dirac equation, approaching the 2QBH via extremal (2+1)QBHs. $\Phi_1$ enters in both $u$ and $v$ \eno{uandv}, but because the shift \eno{Phi1Shift} is constant, only $u$ is affected. Indeed, one may absorb the resulting shift into a redefinition of the zero point of the energy $\omega$ by $g q_1 \Phi_1$,
\eqn{}{
\omega &= \omega_{2+1} + {\sqrt{2} q_1 Q_2 \over L^2} \,, \cr
&= \omega_{2+1} + q_1 \mu_2 \,. 
}
Here $\omega$ is the energy as defined by the Dirac equation that approaches the 2QBH directly, while $\omega_{2+1}$ is the energy in the Dirac equation approaching via extremal (2+1)QBHs.
Since $\mu_1 \to \mu_2$ in this limit, we may interpret the shift as
\eqn{}{
\omega = \omega_{2+1} + q_1 \mu_1 \,,
}
which is precisely the shift in the energy that would accompany a particle of charge $q_1$ with the first chemical potential shifted from $0$ to $\mu_1 = \mu_2$.

One may also interpret this shift in terms of the gap $\Delta$. Given the relations \eno{p1Relation}, \eno{ChargeRelation} (which we will show are necessary for the lift to six dimensions in section~\ref{6DSec}), we have
\eqn{ChargeMagRelation}{
|q_1| = 2 |m_2| \,.
}
Consequently, the shift between the two scales of energy is precisely the magnitude of the gap:
\eqn{OmegaShift}{
\omega = \omega_{2+1} + \sgn(q_1) \Delta \,.
}
In \cite{DeWolfe:2012uv}, many fermions were analyzed in (2+1)QBHs for various values of $0 \leq \mu_R \equiv \mu_1/\mu_2 \leq 1$, at what we are now calling $\omega_{2+1} = 0$. What we have now learned is that the $\mu_R \to 1$ limit of those results should match on to the results found here for the 2QBH. Moreover, $\omega_{2+1} =0$ implies $\omega = \pm \Delta$; thus the ``Fermi surface" results found there match up with the physics not in the middle of the stable region, but at one of its edges. Thus the {\em edge} of the gapped region coincides with the limit of a series of Fermi surfaces.

The gapped region is unusual in the way it appears suddenly for the 2QBH, with energy width $2\Delta$, while if one turns on any nonzero amount of $Q_1$ to give a (2+1)QBH, this region vanishes entirely. The results of this section relate this ``sudden" appearance of the band to the order of limits issue in the parameter space. By choosing how one approaches the 2QBH, one can choose to ``zoom in" on the energy region at the center of the stable region (if $Q_1 =0$ is taken first), or instead one may ``zoom in" on the edge of this band by approaching via the extremal (2+1)QBHs. A limit of $Q_1$, $Q_2$ and $r_H$ that lies in between can presumably be taken to approach some other energy location in the middle of the band.

Given that the Dirac equations coincide, various related quantities should coincide also. We can see this for $\nu^2$. The definition of this quantity in the (2+1)QBHs is \eno{TwoPlusOneNuk}.
We are interested in the $\mu_R \to 1$ limit; expanding we find
\eqn{}{
{\tilde{k} \over \mu_2} \to {k \over \mu_2} -(-1)^\alpha (2 p_1  + \sqrt{2} p_2) = - (-1)^\alpha \left( P + 2 p_1 \right) \,,
}
and thus
\eqn{NuLimit}{
\nu^2 \to {4 m_2^2 - q_1^2  \over 16 (1- \mu_R)} + {1 \over 4} \left( (P + 2 p_1)^2 + 4 m_1 m_2 + {7 q_1^2 \over 8} - {5 m_2^2 \over 2} - {q_1 q_2 \over \sqrt{2}} \right) + {\cal O}(1- \mu_R ) \,.
}
We see there is a leading term that apparently diverges in the limit. However, this term is absent for 
all our fermions, given the relation \eno{ChargeMagRelation}.
Using this relation to eliminate $q_1$ in favor of $\sgn(q_1)$ and $m_2$, we find 
\eqn{}{
\nu^2 \to {1 \over 4} \left( (P + 2 p_1)^2 + 4 m_1 m_2 +m_2^2  - \sgn(q_1 m_2) \sqrt{2} q_2 m_2  \right) \,.
}
The relation \eno{ChargeRelation} also implies a relationship between signs:
\eqn{}{
\sgn(m_2) = - \sgn(q_1) \sgn(p_1) \,.
}
Thanks to \eno{OmegaShift}, we may take $\sgn(\omega) = \sgn(q_1)$ in this limit. Putting these together with \eno{p1Relation},  we have
\eqn{}{
\nu^2 \to  {1 \over 4} \left( (P - {1 \over 2} \sgn(\omega m_2))^2 + 4 m_1 m_2 +m_2^2  - \sgn(\omega m_2) \sqrt{2} q_2 m_2  \right) \,,
}
which exactly matches the expression \eno{NuSquared} for the 2QBH version of $\nu^2$. Thus the dispersion relation for fluctuations has the same scaling in both cases, as it must.

Another quantity providing information about the fluctuation spectrum is $\gamma_k \equiv \arg({\cal G})$. For the (2+1)QBHs, this was determined by \cite{Faulkner:2009wj} to be
\eqn{}{
\gamma_k \equiv \arg \left( \Gamma(-2 \nu_k) \left( e^{-2 \pi i \nu_k} - e^{-2 \pi  (q e)_{\rm eff}} \right) \right)\,,
}
with $(qe)_{\rm eff}^2$ given by the last term in \eno{TwoPlusOneNuk} without the sign:
\eqn{}{
(qe)_{\rm eff}^2= {(\sqrt{2} q_1 \mu_R^3 + q_2 (1 - \mu_R^2) )^2 \over 4 (1 - \mu_R^2)(1 + \mu_R^2)^2} \,.
}
It is easy to see that this goes to infinity as $\mu_R \to 1$ (as long as $q_1 \neq 0$, which holds for all our modes; in fact it is necessarily half-integer quantized by the lift to 6D). For real $\nu_k$ we then have
\eqn{}{
\gamma_k \to - 2 \pi \nu_k \,,
}
which indeed matches the phase  \eno{GammaPhase}.

It is interesting that the limit is only sensible for fermions that can be  lifted to six dimensions. The apparent divergence in \eno{NuLimit} at $\mu_R \to 1$ is related to a term in the Dirac equation: if one takes the limit via extremal (2+1)QBHs with $\omega_{2+1} = 0$, one expects to get the 2QBH equation \eno{TwoQEqn} with $|\omega| = \Delta$, which should have no $r^{-4}$ term and thus should coincide 
with the outer region equation \eno{TwoQOuterFirst}. However, one finds an extra term
\eqn{TwoQOuter}{
 \psi'' +{2 \over r} \psi' +  \left({q_1^2 - 4 m_2^2 \over 2 r^4} + {{1 \over 4} - 4 \nu^2  \over r^2} \right) \psi =0 \,,
}
which vanishes only for fermions obeying \eno{ChargeRelation}. Thus we learn that a generic fermion action, not derived from maximal gauged supergravity, is not consistent throughout the parameter space; approaching the same extremal 2QBHs in two different ways it acquires two inequivalent equations unless \eno{ChargeMagRelation}  is obeyed.

\section{Fermions in ${\cal N}=4$ Super-Yang-Mills}
\label{Nequals4Sec}

We turn now to studying a number of fermionic fluctuations of maximal gauged supergravity, dual to operators in ${\cal N}=4$ Super-Yang-Mills, in the 2QBH backgrounds. The Dirac equations for all spin-1/2 fields were worked out in \cite{DeWolfe:2012uv}.

Here we work out the dispersion relation between $k$ and $\omega$ in the entire stable region for each fermion. The Fermi surface is at $\omega =0$; we find such a surface and the corresponding $k_F$, and then vary $\omega$ away from zero and find the $k$ giving a pole in the Green's function in each case. Close to the Fermi surface pole the dispersion is linear, and primarily deviates from this behavior in the vicinity of $|\omega| = \Delta$.
At $|\omega| = \Delta$, the results match on to the $\mu_R \to 1$ limit of the study of (2+1)QBHs at $\omega_{2+1} = 0$ in \cite{DeWolfe:2012uv}, as described in the last section.

In addition to the poles connected to the Fermi surface singularity, we also find other pairs of poles nucleating at nonzero $\omega$. These additional fluctuations always seem to appear near an oscillatory region living at the gap, and generically the poles fall into the region and cease to exist. In a few cases, however, they miss the oscillatory region and survive as excitations at the gap, which also match with Fermi surfaces seen in the (2+1)QBHs in \cite{DeWolfe:2012uv}.

\subsection{Fermion A}

\begin{figure}
\begin{center}
\includegraphics[scale=0.8]{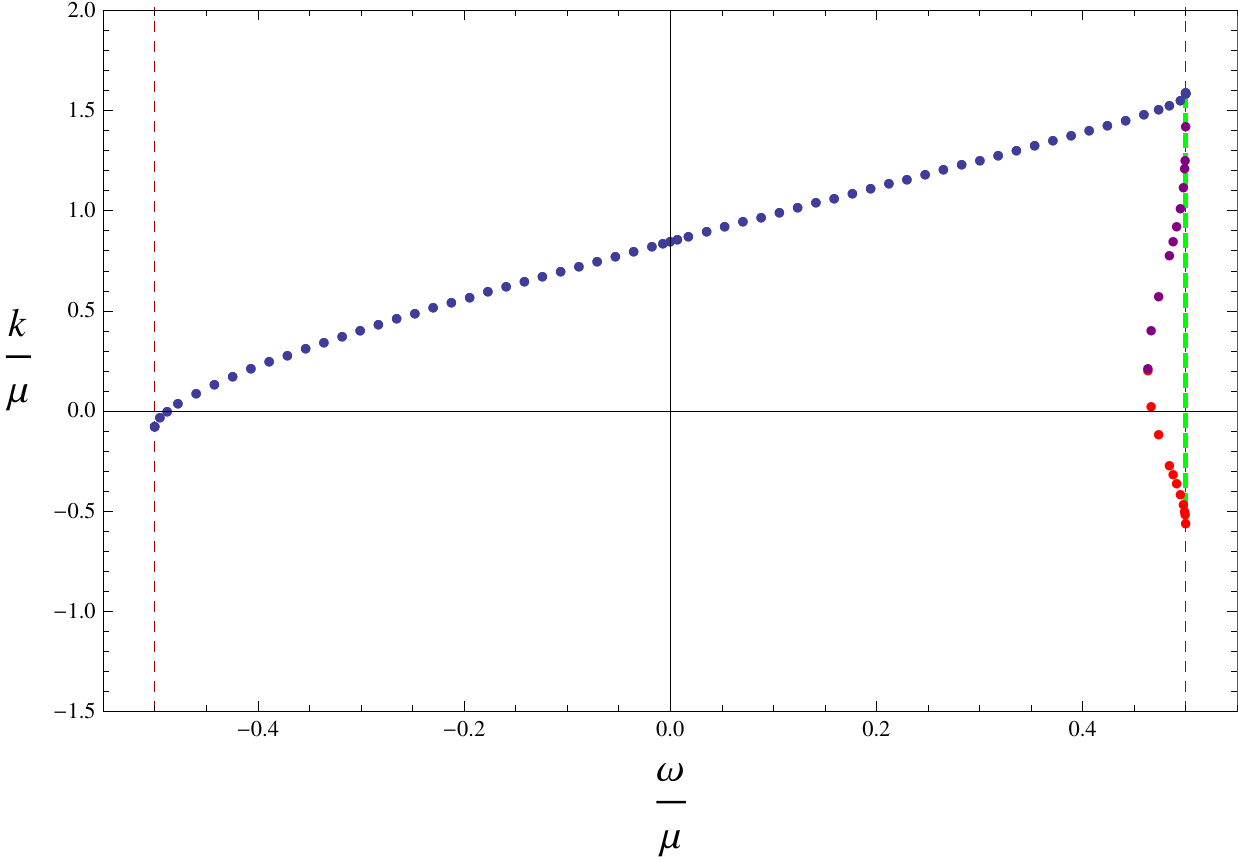}
\caption{ Poles in the retarded Green's function from $\omega =0 $ to $\omega = \pm \Delta$ (red dashed lines) for fermion A.
\label{DispersionAFig}}
\end{center}
\end{figure}

Consider first the pair of distinct fermions \cite{DeWolfe:2012uv}

s\begin{equation}
\label{FermionATable}
\begin{tabular}{|c|c|c|c|c|c|c|c|} \hline
$\chi^{q_a q_b q_c}$ &Dual operator &$m_1$& $m_2$ & $q_1$& $q_2$  & $p_1$ & $p_2$\\ \hline
$\chi^{({1 \over 2}, {3 \over 2}, {1 \over 2})}$, $\chi^{({1 \over 2}, { 1\over 2}, {3 \over 2})}$ & $\lambda_1 Z_2$, $\lambda_1 Z_3$&${1 \over 2}$& $-{1 \over 4}$  & ${1 \over 2}$& $2$ & ${1 \over 4} $& $0$\\
$\bar\chi^{(-{1 \over 2},  {3 \over 2}, {1 \over 2})}$, $\bar\chi^{(-{1 \over 2},  {1 \over 2}, {3 \over 2})}$ & $\overline\lambda_2 Z_2$, $\overline\lambda_2 Z_3$&${1 \over 2}$& $-{1 \over 4}$ & $-{1 \over 2}$& $2$ & $-{1 \over 4}$& $0$ \\ \hline
\end{tabular}
\end{equation}
In the 2QBH background, where $q_1$ and $p_1$ are irrelevant, these fermions have the same Dirac equation with $(m_1, m_2) = (1/2, -1/4)$ and $(q_2, p_2) = (2,0)$.
In this case $\Delta$ is just half the chemical potential,
\eqn{}{
\Delta \equiv 2 |m_2| \mu = {\mu \over 2} \,.
}
It was found in \cite{DeWolfe:2012uv} that there is a Fermi surface (imposing the regular boundary condition) at
\eqn{}{
{k_F \over \mu} \approx 0.83934 \,.
}
Continuing to use the regular boundary condition, we can numerically solve for poles in the Green's function from $\omega =0$ up to $\omega = \pm \Delta$; in so doing $k$ varies continuously from $k_F$ up to $k_{\pm\Delta}$. The result of this procedure is depicted as the blue dots in figure~\ref{DispersionAFig}. The dispersion is approximately linear at small values of $\omega/\mu$, with the form
\eqn{}{
\omega \approx v_F\,k_{\perp}\,, \quad \quad v_F \approx 0.724 \,,
}
which is about 5/4 times larger than the speed of sound $c_s =1/\sqrt{3}$ in this (conformal) background.
At $\omega = -\Delta$, we have
\eqn{}{
{k_{-\Delta} - k_F \over \mu} \approx  -0.92513\,,
}
with 
\eqn{}{
\nu_{k_{-\Delta}} = 0.33211 \,,
}
associated to a non-Fermi liquid.
 For $\omega = \Delta$
\eqn{}{
{k_\Delta - k_F \over \mu} \approx 0.743052\,,
}
implying
\eqn{}{
{k_\Delta \over \mu} =  1.58239 \,,
}
at which point
\eqn{FermionARightNu}{
\nu_{k_\Delta} = 0.08211 \,.
}
For fluctuations above $\omega = \Delta$, we again have a non-Fermi liquid. Outside the stable region, $\omega$ will develop an imaginary part as well, and move off into the complex plane. 

\begin{figure}
\begin{center}
\includegraphics[scale=0.9]{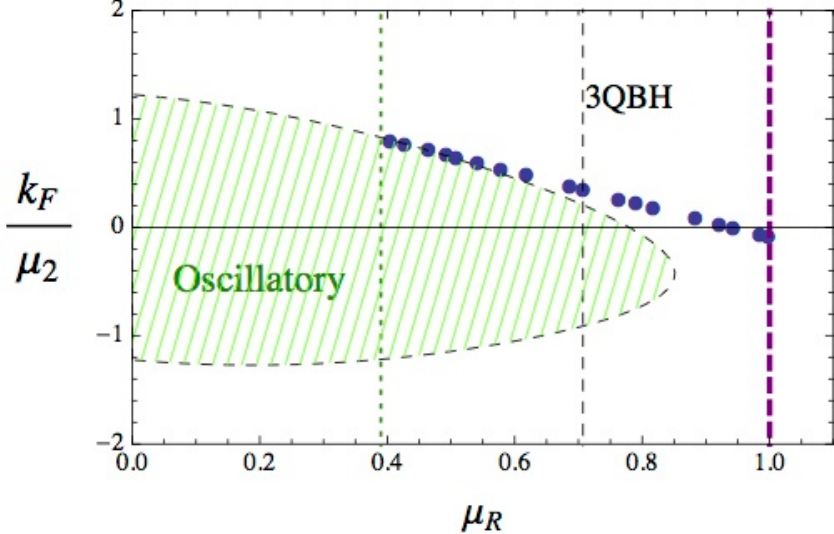}
\includegraphics[scale=0.9]{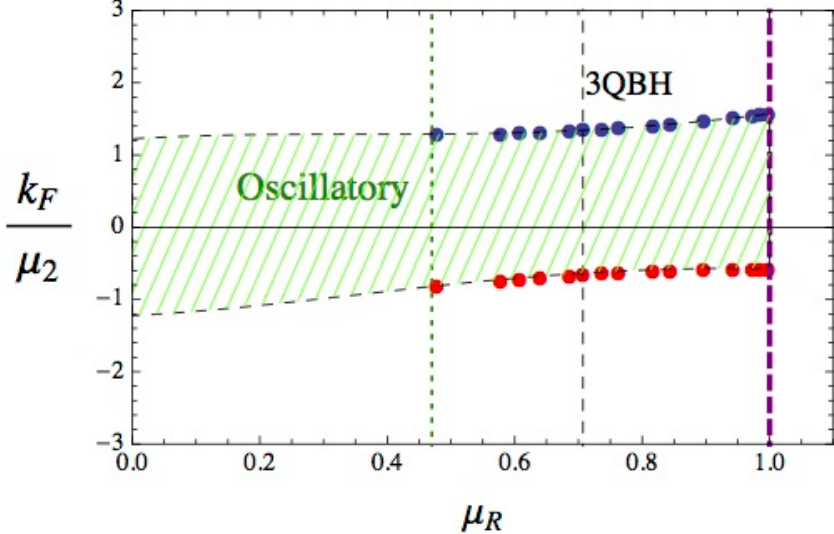}
\caption{Fermi surfaces in (2+1)-charge black holes as a function of $\mu_R$ from \cite{DeWolfe:2012uv}; on the left is case 3, on the right case 1. The $\mu_R \to 1$ limits (purple dashed lines) match onto the left and right edges of the gapped region in figure~\ref{DispersionAFig}.}
\label{Cases31Fig}
\end{center}
\end{figure}

The behavior at $\omega = \pm \Delta$ should match on to results from the extremal (2+1)QBHs found in \cite{DeWolfe:2012uv}. Here the two fermions in \eno{FermionATable} behave differently according to \eno{OmegaShift} since they have opposite signs of $q_1$: the first line in the table, called case 1 in \cite{DeWolfe:2012uv}, has $\omega = \omega_{2+1} + \Delta$ and thus the extremal (2+1)QBH limit ``zooms in" on the $\omega = \Delta$ side of the gapped region, while the second line in the table, case 3 in \cite{DeWolfe:2012uv}, zooms in on the $\omega= - \Delta$ side.

We reproduce the plots from \cite{DeWolfe:2012uv} showing locations of Fermi surfaces at various values of $0 \leq \mu_R \leq 1$ for cases 3 and 1 in figure~\ref{Cases31Fig}. The values of $k_F$ at the $\mu_R \to 1$ limit must match to  values of $k_{-\Delta}$ and $k_\Delta$ in figure~\ref{DispersionAFig}, respectively, and indeed the Fermi surfaces indicated by blue dots do in each case. Values of $\nu_k$ match as well.

There is a new feature here, however: case 1 has two distinct Fermi surfaces, one on each side of an oscillatory region. (Since the oscillatory region is defined by $\nu^2 < 0$, and $\nu$ necessarily matches between the two limits, the oscillatory region is also present in the 2QBH case, and is indicated by a green line in figure~\ref{DispersionAFig}.) Thus there ought to be another pole in the 2QBH Green's function at $\omega = \Delta$, one not connected to the pole at $\omega =0$.

Indeed we find this is the case. Near $\omega = 0$, the only pole in the Green's function is the one connected to the Fermi surface pole at $\omega =0$. But at a certain value $0 < \omega < \Delta$, a new pair of poles nucleates. As $\omega$ is increased further, they spread apart with different values of $k$. One pole (indicated in red in figure~\ref{DispersionAFig}) reaches the $\omega = \Delta$ line just below the oscillatory region, and matches the red pole from figure~\ref{Cases31Fig}. This pole has 
\eqn{}{
{k_\Delta \over \mu} \approx  -0.58239 \,,
}
and the same value of  $\nu_{k_\Delta}$  as the other pole at $\omega = \Delta$ \eno{FermionARightNu}. Meanwhile, the partner pole, indicated in figure~\ref{DispersionAFig} in purple, reaches the $\omega = \Delta$ axis inside the oscillatory region, and does not survive. Thus the overall picture is consistent between the (2+1)QBHs from \cite{DeWolfe:2012uv} and the current work.

As one moves even closer to $\omega = \Delta$, several further pairs of poles nucleate, all of which disappear into the oscillatory region; these continue to appear to the limits of our numerical precision. These pairs lie very close to $\omega = \Delta$ and are not indicated in the figure. Thus there appears to be an accumulation of additional states near the oscillatory region.

\subsection{Fermion B}

\begin{figure}
\begin{center}
\includegraphics[scale=0.8]{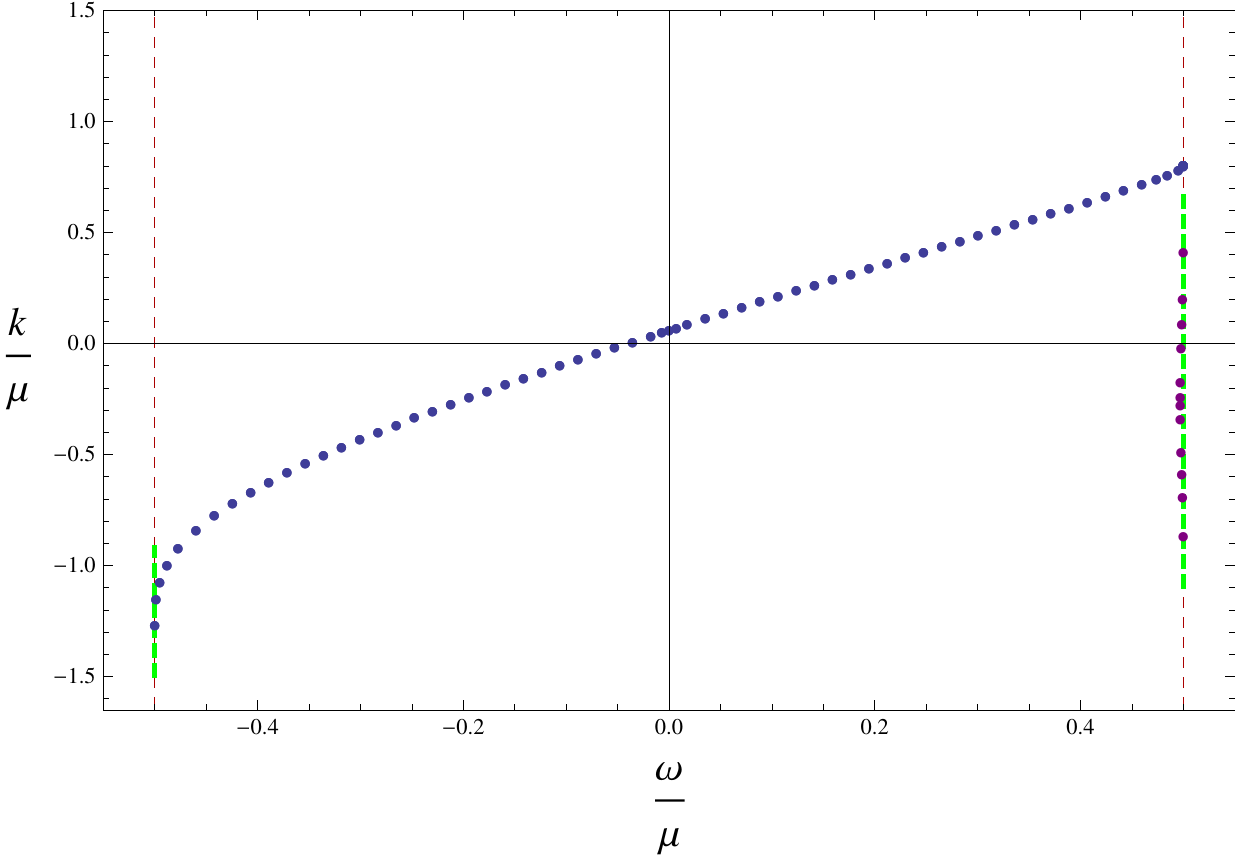}
\caption{ Poles in the retarded Green's function from $\omega =0 $ to $\omega = \pm \Delta$ (red dashed lines) for fermion B.}
\label{DispersionBFig}
\end{center}
\end{figure}

Consider now a second pair of fermions whose charges coincide when $q_1$, $p_1$ are neglected:
\begin{equation}
\label{FermionBTable}
\begin{tabular}{|c|c|c|c|c|c|c|c|} \hline
$\chi^{q_a q_b q_c}$ &Dual operator &$m_1$& $m_2$ & $q_1$& $q_2$  & $p_1$ & $p_2$\\ \hline
$\chi^{(-{1 \over 2},  {3 \over 2}, -{1 \over 2})}$, $\chi^{(-{1 \over 2}, -{1 \over 2},  {3 \over 2})}$& $\lambda_3 Z_2$, $\lambda_4 Z_3$&${1 \over 2}$& $-{1 \over 4}$& $-{1 \over 2}$& $1$ & $- {1 \over 4}$& $- {1 \over 2}$\\
$\bar\chi^{({1 \over 2}, -{1 \over 2},  {3 \over 2})}$, $\bar\chi^{({1 \over 2},  {3 \over 2}, -{1 \over 2})}$& $\overline\lambda_3 Z_3$, $\overline\lambda_4 Z_2$& ${1 \over 2}$& $-{1 \over 4}$ & ${1 \over 2}$& $1$ & ${1 \over 4}$& $-{1 \over 2}$ \\ \hline
\end{tabular}
\end{equation}
These behave the same for the 2QBH and in \cite{DeWolfe:2012uv} a Fermi surface was found at
\eqn{}{
\frac{k_F}{\mu} \approx 0.05202 \,.
}
Again $\Delta = \mu/2$, we can follow the pole away from $\omega = 0$ all the way to $\omega= \pm \Delta$;  this is plotted in figure~\ref{DispersionBFig}. As with fermion A, the dispersion at small $\omega/\mu$ is linear, and the corresponding Fermi velocity $v_F\approx 0.678$ is similar. At $\omega = \Delta$, the pole hits the axis at
\eqn{}{
{k_\Delta \over \mu} \approx 0.79289 \,,
}
resulting in 
\eqn{}{
\nu_{k_\Delta} = 0.22855 \,,
}
yet another non-Fermi liquid. Approaching $\omega = -\Delta$, however, the line of poles reaches the axis inside an oscillatory region, and there is no solution at $\omega = -\Delta$ precisely.

\begin{figure}
\begin{center}
\includegraphics[scale=0.9]{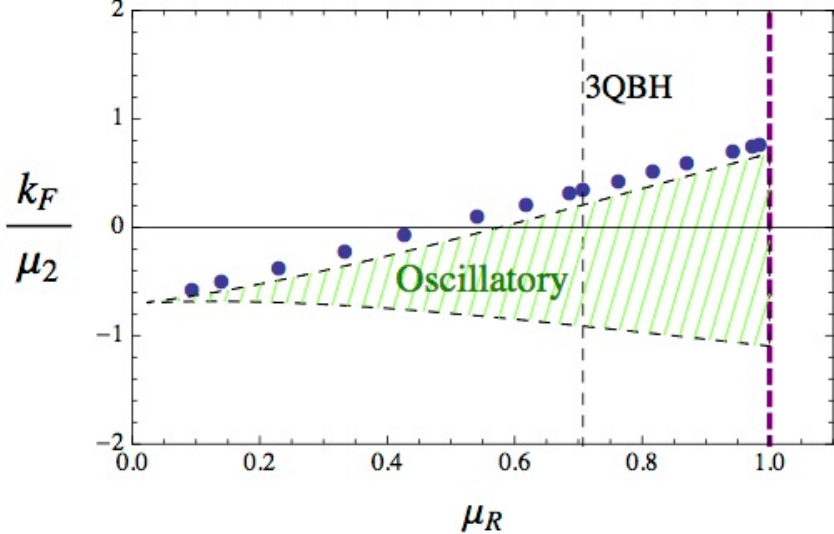}
\includegraphics[scale=0.9]{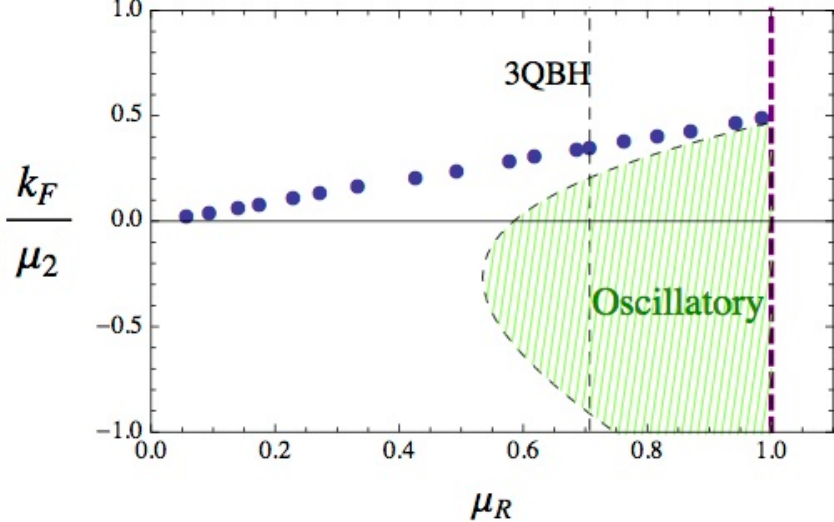}
\caption{Fermi surfaces in (2+1)-charge black holes as a function of $\mu_R$ from \cite{DeWolfe:2012uv} for cases  4 and 5. The $\mu_R \to 1$ limit (purple dashed lines) matches onto the right edges of the gapped region in figures~\ref{DispersionBFig} and \ref{DispersionNeutralFig}, respectively.}
\label{Case4Fig}
\end{center}
\end{figure}

Again we expect a match to results from \cite{DeWolfe:2012uv} for (2+1)QBHs. The first fermion listed in the table \eno{FermionBTable} has $q_1 < 0$, and hence zooms in on $\omega = - \Delta$, where we found no pole. Indeed, no Fermi surfaces were found for this fermion in \cite{DeWolfe:2012uv}. For the second fermion in the table, $q_1> 0$ and it zooms in on $\omega = \Delta$. This fermion was called case 4 in \cite{DeWolfe:2012uv}, and its Fermi surfaces are reproduced in figure~\ref{Case4Fig}. We again see that the value of $k_F/\mu$ at $\mu_R \to 1$ matches the pole at $\omega =\Delta$ in figure~\ref{DispersionBFig}.

There are oscillatory regions on both sides of the gap $\omega = \pm \Delta$. Again we find a proliferation of additional poles near the $\omega= \Delta$ oscillatory region; the first such pair is plotted in figure~\ref{DispersionBFig} and lies almost on top of the region. No such poles are found for the $\omega=  -\Delta$ oscillatory region.

\subsection{Neutral fermion}

\begin{figure}
\begin{center}
\includegraphics[scale=0.8]{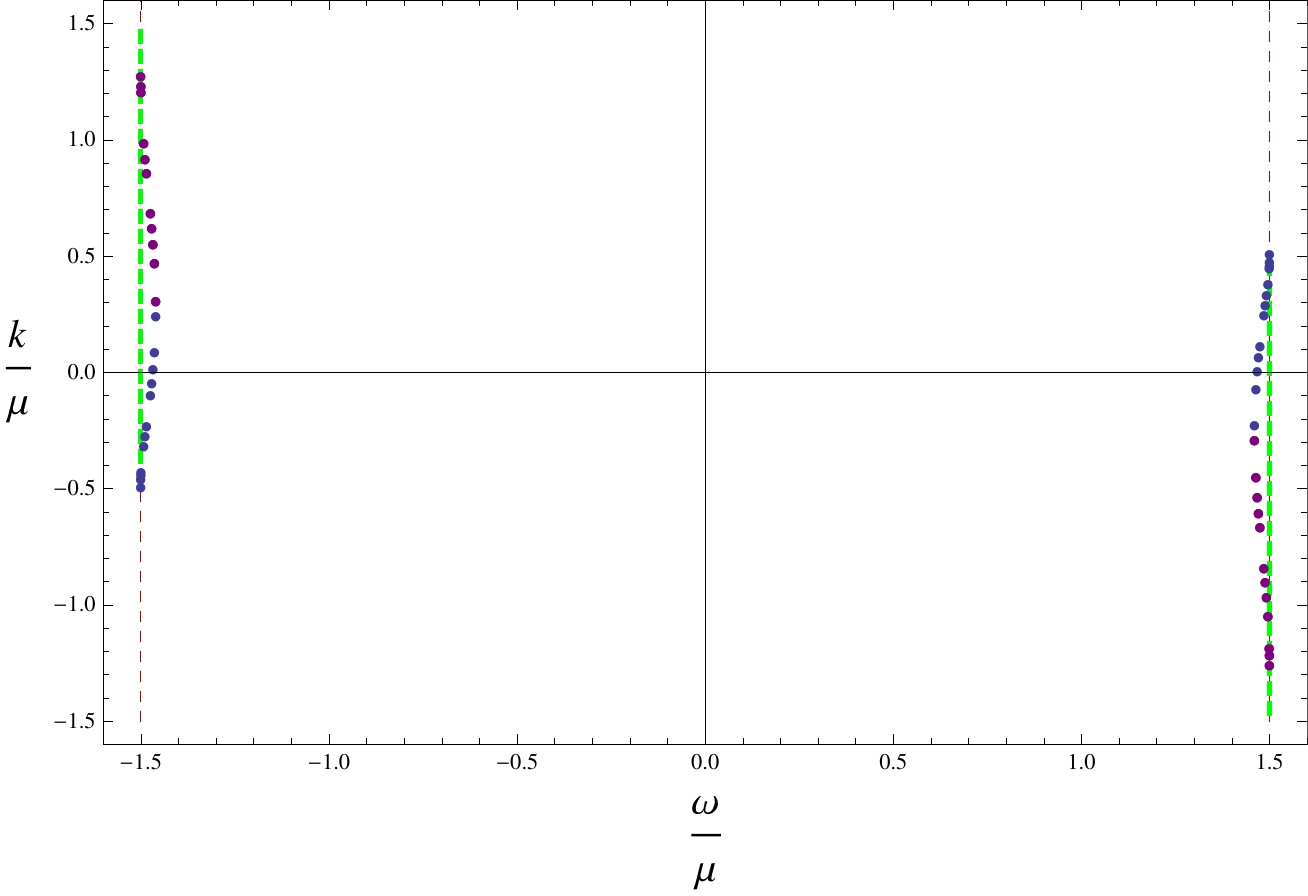}
\caption{ Poles in the retarded Green's function from $\omega =0 $ to $\omega = \pm \Delta$ (red dashed lines) for the neutral fermion.}
\label{DispersionNeutralFig}
\end{center}
\end{figure}

One may also consider the fermion
\begin{equation}
\begin{tabular}{|c|c|c|c|c|c|c|c|} \hline
$\chi^{q_a q_b q_c}$ &Dual operator &$m_1$& $m_2$ & $q_1$& $q_2$  & $p_1$ & $p_2$\\ \hline
$\bar\chi^{({3 \over 2}, -{1 \over 2}, {1 \over 2})}$  , $\bar\chi^{({3 \over 2}, {1 \over 2}, -{1 \over 2})}$ & $\overline\lambda_3 Z_1$, $\overline\lambda_4 Z_1$&$-{1 \over 2}$& ${3 \over 4}$  & ${3 \over 2}$& $0$ & $-{1 \over 4} $&$0$\\ \hline
\end{tabular}
\end{equation}
called case 5 in \cite{DeWolfe:2012uv}. This case is neutral as far as the 2QBH is concerned, with $\Delta = 3\mu/2$, and no Fermi surface was found at $\omega = 0$. However, the results of  \cite{DeWolfe:2012uv} shown in figure~\ref{Case4Fig} indicate there should be a singularity at $\omega = \Delta$ to match the Fermi surface at $\mu_R \to 1$.
This can occur if a pair of singularities nucleate at some nonzero $\omega$, one makes it to the edge of the gap while the other falls into the oscillatory region. This is indeed what we find  in figure~\ref{DispersionNeutralFig}, where the lower pole falls into the oscillatory region  while the upper pole escapes, matching the pole in figure~\ref{Case4Fig}. Due to the vanishing of all gauge couplings for this case, the Dirac equation is symmetric under $(k, \omega) \to (-k, -\omega)$, and we find identical behavior at $\omega = -\Delta$.

The sole remaining fermion for which a Fermi surface was found in \cite{DeWolfe:2012uv}, there called case 2, was characterized by the charges
\begin{equation}
\begin{tabular}{|c|c|c|c|c|c|c|c|} \hline
$\chi^{q_a q_b q_c}$ &Dual operator &$m_1$& $m_2$ & $q_1$& $q_2$  & $p_1$ & $p_2$\\ \hline
$\chi^{({3 \over 2}, {1 \over 2}, {1 \over 2})}$ &$\lambda_1 Z_1$ &$-{1 \over 2}$& ${3 \over 4}$ & ${3 \over 2}$& $1$ & $- {1 \over 4}$& ${1 \over 2}$\\ \hline
\end{tabular}
\end{equation}
While this fermion has Fermi surface singularities for many values of $\mu_R$, the line of Fermi surfaces falls into the oscillatory region before reaching $\mu_R = 1$. Thus we expect no singularity at $\omega = \Delta$ for this case; in fact we find no Fermi surface at $\omega = 0$, either.

\section{Six-dimensional lift}
\label{6DSec}

One concern about the 2QBH is its singularity at the horizon. One may also wish to understand the geometrical nature of the inner region a little better. Here we show that in the near-horizon limit, the geometry can be lifted to $AdS_3 \times \mathbb{R}^3$. Moreover, the fermions lift to  six-dimensional fermions as well; this lift identifies the charge $q_1$ as Kaluza-Klein momentum in the sixth direction, and places relations amongst the parameters $q_1$, $m_2$ and $p_1$.

The six-dimensional lift we demonstrate cannot be considered the full picture, as several aspects of the fermion are subleading (in the near-horizon expansion) and do not appear, in particular $m_1$ and $q_2$. Presumably including more subleading terms would provide a complete picture, ultimately recovering the intermediate compactification between the (squashed) $AdS_5 \times S^5$ lift and the five-dimensional reduction. The lift we display, however, elegantly resolves the singularity and establishes the parameter relations, and so is sufficient for our purposes.

\subsection{Kaluza-Klein Oxidation of Metric}

The near-horizon ($r \to 0$) limit of the 5D extremal  2QBH metric is
\eqn{}{
ds^2 = - {2 r^{8/3} \over L^2 Q^{2/3}} dt^2 + {r^{2/3} Q^{4/3} \over L^2} d\vec{x}^2 + {L^2 \over 2 r^{4/3} Q^{2/3}} dr^2 \,,
}
and the scalar diverges as
\eqn{}{
e^{{\phi \over 2 \sqrt{6}}} = \left( Q \over r\right)^{1/3} \,.
}
We now claim that this metric can be obtained as the dimensional reduction of a 6D metric.  Using a hat for the 6D metric, the Kaluza-Klein reduction ansatz is
\eqn{}{
d\hat{s}^2 &= e^{2 \alpha \phi} ds^2 + e^{2 \beta \phi} (dz + {\cal A})^2 \cr
&=e^{\phi \over \sqrt{6} }ds^2 + e^{- {3 \phi \over \sqrt{6}}} (L d \varphi_3 + {\cal A})^2 \,,
}
where  we defined $z \equiv L \varphi_3$, and
\eqn{}{
\alpha = {1 \over 2 \sqrt{6}}\,, \quad \quad\beta = -3 \alpha =- {3 \over 2 \sqrt{6}} \,.
} 
The 6D metric is then AdS$_3 \times \mathbb{R}^3$:
\eqn{}{
d\hat{s}^2 = -{2 r^2\over L^2} dt^2 + {L^2 \over 2 r^2} dr^2 + {r^2 L^2 \over Q^2} d \varphi_3^2 + {Q^2 \over L^2} d\vec{x}^2 \,.
}
One can calculate the sechsbeins and spin connections; we present them in Appendix C.
The dimensional reduction of the Einstein-Hilbert action then gives
\eqn{}{
{\cal L}_{EH} = \sqrt{-\hat{g}} \hat{R} = \sqrt{-g} \left( R - {1 \over 2} (\partial \phi)^2 - {1 \over 4} e^{-4 \phi \over \sqrt{6}} {\cal F}^2 \right) \,,
}
and the action for the KK gauge field matches that for the 1Q gauge field $a_\mu$ (the one not turned on in the 2QBH background) if we define
\eqn{}{
a_\mu \equiv {1 \over 2} {\cal A}_\mu \,.
}
Thus the charge $q_1$ is identified with Kaluza-Klein momentum.

We may also obtain the leading term in the five-dimensional potential as a reduction of a six-dimensional cosmological constant.  Consider a cosmological term in 6D,
\eqn{}{
{\cal L}_\Lambda = \sqrt{-\hat{g}} \hat\Lambda \,,
}
and dimensionally reduce it, leading to
\eqn{}{
{\cal L}_\Lambda = \sqrt{-g} e^{\phi \over \sqrt{6}} \hat\Lambda\,.
}
The potential terms in the 5D theory are
\eqn{}{
{\cal L}_{\rm pot} = \sqrt{-g}\left({8 \over L^2} e^{\phi \over \sqrt{6}} + {4 \over L^2} e^{-2\phi\over \sqrt{6}} \right) \,,
}
and the first term will dominate near the horizon. We can match the 6D cosmological constant to the leading term if we choose
\eqn{}{
\hat\Lambda = {8 \over L^2} \,,
}
and in 6D we have the term
\eqn{}{
{\cal L}_\Lambda = \sqrt{-\hat{g}} {8 \over L^2} \,.
}
Note that there is no obvious way to get the subleading term in this limit, since in 6D $\phi$ is part of the metric and one cannot simply write down terms with arbitrary powers of it.

\subsection{Oxidation of gauge field}

The gauge field $a_\mu$ is the 6D graviphoton; the other 5D gauge field $A_\mu$ must have another source in six dimensions. In six dimensions, one can consider a 1-form gauge field or a 2-form gauge field.  For the one-form case,
the reduction from six dimensions to five is
\eqn{}{
- {1 \over 2} \hat{F}_2 \wedge \hat{*} \hat{F}_2 = - {1 \over 2}e^{-2 \alpha \phi} F_2 \wedge * F_2
- {1 \over 2}e^{6 \alpha \phi} F_1 \wedge * F_1 \,,
}
where
\eqn{}{
F_2 \equiv d A_1 - dA_0 \wedge {\cal A}_1 \,, \quad \quad F_1 = dA_0 \,.
}
Meanwhile, reducing a 6D 2-form potential we have
\eqn{}{
- {1 \over 2} \hat{H}_3 \wedge \hat{*} \hat{H}_3 = - {1 \over 2}e^{-4 \alpha \phi} H_3 \wedge * H_3
- {1 \over 2}e^{4 \alpha \phi} H_2 \wedge * H_2 \,,
}
where
\eqn{}{
H_3 \equiv d B_2 - dB_1 \wedge {\cal A}_1 \,, \quad \quad H_2 = dB_1 \,.
}
Thus we see that in order to match the gauge field $A_\mu$ in five-dimensions, we need to reduce a 6D two-form, and identify
\eqn{}{
B_\mu \propto A_\mu \,.
}
Details of the coefficient of the kinetic term depend on whether we take $H_3$ to be self-dual or not; both matching the 5D Chern-Simons term, and the presumed ten-dimensional origin of this $H_3$ in the five-form $F_5$, suggest that it should be. The normalization will not be needed for the arguments we make.

\subsection{Probe scalar reduction and interpretation of gap $\Delta$}

Although we are ultimately interested in fermions, it is edifying to consider the simpler case of a 6D complex scalar $\hat\eta$ first. Starting with the six-dimensional action
\eqn{}{
S_\eta \equiv -  \int d^6x \sqrt{-\hat{g}} \hat{g}^{MN} \partial_M \hat\eta \partial_N \hat\eta \,,
}
we assume the KK ansatz
\eqn{}{
\hat\eta(\hat{x}^M) = e^{i n \varphi_3} \eta(x^\mu) \,,
}
and arrive at the five-dimensional action
\eqn{}{
S_\eta \to - 2 \pi L \int d^5x \sqrt{-g} \, \left(g^{\mu\nu} \overline{D_\mu \eta} D_\nu \eta + e^{4 \phi \over \sqrt{6}} {n^2 \over L^2} |\eta|^2 \right) \,,
}
where the covariant derivative is
\eqn{}{
D_\mu \equiv \partial_\mu - i {n \over L} {\cal A}_\mu = \partial_\mu - i g n a_\mu \,,
}
having used $g \equiv 2/L$. We see that the KK momentum $n$ is also the charge $q_1$. Moreover there is an effective mass,
\eqn{}{
m = \pm {n \over L} e^{2 \phi \over \sqrt{6}} \,.
}
This matches the leading term in the mass function \eno{MassFunction} provided $2 m_2 = \pm q_1$. This relation will hold for the fermions as well.

With the relation between $m_2$ and the KK momentum $n$ in hand, we can provide a new interpretation of the gap $\Delta$. Consider a six-dimensional momentum vector in the KK/time directions $k_M \equiv (k_{\varphi_3}, k_t, 0, 0, 0, 0) = (n, \omega, 0, 0, 0, 0)$. The norm of this vector is
\eqn{}{
k^M k_M &= {Q^2 \over r^2 L^2} n^2 - {L^2 \over 2 r^2} \omega^2 \,, \cr
&= {L^2 \over 2 r^2} (\Delta^2 - \omega^2)  \,,
}
where we used $n^2 = 2 m_2^2$ and that $\Delta^2 = 8 m_2^2 Q^2/L^4$. Thus the gap represents the minimum energy required for the six-dimensional momentum to be timelike; fluctuations of lesser energy, in the gapped region, correspond to spacelike 6D momenta due to the contribution of the momentum in the compact direction.

The emergence of an $AdS_3$ region suggests aspects of the IR physics are controlled by a dual CFT$_2$. A further understanding of the associated non-chiral Virasoro algebra could lead to greater insight concerning the nature of the gap.

\subsection{Fermion reduction}

Turn now to the fermionic case. The five-dimensional spinors in maximal gauged supergravity are four-component symplectic Majorana pairs. These descend from eight-component Majorana-Weyl fermions $\lambda^a$ in six dimensions:
\eqn{}{
\Gamma \lambda^a  =  \varepsilon\lambda^a \,, \quad \quad \lambda^a = B_6 \Omega_{ab} (\lambda^b)^*\,.
}
Here $\varepsilon = \pm 1$ gives the chirality, $B_6$ is the six-dimensional conjugation matrix, and $\Omega_{ab}$ is an antisymmetric matrix in the space of spinors; details are relegated to Appendix B. 
The kinetic term in the 6D Lagrangian is 
\eqn{Kinetic6D}{
{\cal L} = \sqrt{-\hat{g}} {i \over 2} \bar\lambda^a \Gamma^{\underline{M}} \,\hat{e}^N_{\;\;\;\underline{M}} \hat\nabla_N \lambda_a \,,
}
and a mass term is forbidden by the Weyl condition.  The covariant derivative is
\eqn{}{
\hat\nabla_N = \partial_N - {1 \over 4} \hat\omega_N^{\;\;\;\;\underline{PQ}} \Gamma_{\underline{PQ}} \,.
}
Let us reduce the kinetic term to five dimensions.
Because of the symplectic Majorana condition, if we take one 6D spinor to depend on $\varphi_3 \equiv z/L$ as a complex exponential, the other must depend on the conjugate.  Furthermore, we will allow a rescaling by a power of the scalar $\phi$. We choose a basis for $\Gamma$-matrices where the Weyl condition for one chirality is solved by spinors of the form
\eqn{FermionReductionAnsatz}{
\lambda_1(z, x^\mu) \equiv e^{i n \varphi_3} e^{{\eta  \over \sqrt{6}}\phi} \begin{pmatrix}\chi_1(x^\mu) \cr 0 \end{pmatrix}  \,, \quad \quad 
\lambda_2(z, x^\mu) \equiv e^{-i n \varphi_3} e^{{\eta  \over \sqrt{6}}\phi} \begin{pmatrix} \chi_2(x^\mu) \cr 0\end{pmatrix} \,,
}
where the $\chi_a$ are 4-component spinors that are symplectic Majorana in a 5D sense, and $0$ stands for vanishing 4-component spinors. The other chirality would have the $\chi_i$ below the $0$.
Here $n$ is half-integer quantized since the fields are fermionic.

Performing the reduction (see Appendix C), we find we can obtain a canonical 5D kinetic term with the choice $\eta = -1/4$. Our total 5D Lagrangian becomes
\eqn{FinalLagrangian}{
e^{-1} {\cal L} =  {i\over2} \bar\chi \gamma^\mu \nabla_\mu \chi + {1 \over 2}\bar\chi \left(n \varepsilon \over L \right)e^{2 \phi \over \sqrt{6}} \tau_3 \chi + {1 \over 2} \bar\chi \left( 2 n \over L\right) \gamma^\mu a_\mu \tau_3 \chi 
+ {i \over 2} \bar\chi \left( \varepsilon \over 4 \right) e^{-2 \phi \over \sqrt{6}} \gamma^{\mu\nu} f_{\mu\nu} \chi  \,,
}
where the $\tau_3$ Pauli matrix acts in the the symplectic Majorana space (details in the appendices). 
Comparing to the canonical form of the Lagrangian, 
\eqn{}{
{\cal L} = {1\over 2} \left(i \bar\chi \gamma^\mu \nabla_\mu \chi - m  \bar\chi \tau_3 \chi + q \bar\chi \gamma^\mu a_\mu \tau_3 \chi + ip \bar\chi \gamma^{\mu\nu} f_{\mu\nu} \chi \right) \,,
}
and using that in the 2QBH we have masses of the form
\eqn{}{
m(\phi) = g \left( m_1 e^{-\phi \over \sqrt{6}} + m_2 e^{2\phi \over \sqrt{6}} \right) \approx {2 m_2 \over L} e^{2 \phi \over \sqrt{6}} + \ldots \,,
}
we identify an effective mass, gauge coupling and Pauli coupling,

\eqn{}{
m_2 = - {n \varepsilon \over 2} \,, \quad \quad
q_1 =   n \,, \quad \quad
p_1 = {\varepsilon \over 4} \,.
}
Thus these three five-dimensional quantities are determined by two six-dimensional ones: the KK momentum $n$, and the chirality $\varepsilon$.  Since $n$ is half-integer quantized, we expect $q_1$ to be likewise, and $m_2$ to be quarter-integer quantized; we also find  $|p_1| = 1/4$.  Since three quantities are determined from two underlying parameters, we have a relation, which we can express as
\eqn{CRagain}{
m_2 = - 2 q_1 p_1 \,,
}
which is a result we advertised previously as equation \eno{ChargeRelation}.

Let us now compare these results to the fermions we know in five dimensions.
Below is the table from \cite{DeWolfe:2012uv}.\footnote{In \cite{DeWolfe:2012uv}, for all the fields whose near-boundary mass came out negative (indicated as $\bar\chi$ in the first column), we flipped the signs of the gamma-matrix basis, thus flipping the signs of $m_i$ and $p_i$.  We have undone those flips here.} One may verify that all the implied properties are present: $q_1$ is half-integer quantized ($|q_1| = 1/2$ or $|q_1| = 3/2$), $m_2$ is correspondingly quarter integer quantized, $|p_1| = 1/4$, and the relation \eno{ChargeRelation} is satisfied.  One can thus derive the associated KK charge $n$ and 6D chirality $\varepsilon$, and they are added in the last two columns of the table.  We find various modes involving the two lowest half-integer quantized charges, as well as both chiralities:
\begin{equation}
\begin{tabular}{|c|c|c|c|c|c|c|c||c|c|} \hline
$\chi^{q_a q_b q_c}$ &Dual operator &$m_1$& $m_2$ & $q_1$& $q_2$  & $p_1$ & $p_2$& $n$ &$\varepsilon$ \\ \hline
$\chi^{({3 \over 2}, {1 \over 2}, {1 \over 2})}$ &$\lambda_1 Z_1$ &$-{1 \over 2}$& ${3 \over 4}$ & ${3 \over 2}$& $1$ & $- {1 \over 4}$& ${1 \over 2}$ &${3 \over 2}$&$-1$\\
$\chi^{({3 \over 2}, -{1 \over 2}, -{1 \over 2})}$ & $\lambda_2 Z_1$&$-{1 \over 2}$& ${3 \over 4}$  & ${3 \over 2}$& $-1$ & $- {1 \over 4} $& $-{1 \over 2}$&${3 \over 2}$&$-1$\\
$\bar\chi^{({3 \over 2}, -{1 \over 2}, {1 \over 2})}$  , $\bar\chi^{({3 \over 2}, {1 \over 2}, -{1 \over 2})}$ & $\overline\lambda_3 Z_1$, $\overline\lambda_4 Z_1$&${1 \over 2}$& $-{3 \over 4}$  & ${3 \over 2}$& $0$ & ${1 \over 4} $&$0$&${3 \over 2}$&$1$\\ \hline
$\chi^{({1 \over 2}, {3 \over 2}, {1 \over 2})}$, $\chi^{({1 \over 2}, { 1\over 2}, {3 \over 2})}$ & $\lambda_1 Z_2$, $\lambda_1 Z_3$&${1 \over 2}$& $-{1 \over 4}$  & ${1 \over 2}$& $2$ & ${1 \over 4} $& $0$&${1 \over 2}$&$1$\\
$\bar\chi^{(-{1 \over 2},  {3 \over 2}, {1 \over 2})}$, $\bar\chi^{(-{1 \over 2},  {1 \over 2}, {3 \over 2})}$ & $\overline\lambda_2 Z_2$, $\overline\lambda_2 Z_3$&$-{1 \over 2}$& ${1 \over 4}$ & $-{1 \over 2}$& $2$ & ${1 \over 4}$& $0$ &$-{1 \over 2}$&$1$\\ 
$\chi^{(-{1 \over 2},  {3 \over 2}, -{1 \over 2})}$, $\chi^{(-{1 \over 2}, -{1 \over 2},  {3 \over 2})}$& $\lambda_3 Z_2$, $\lambda_4 Z_3$&${1 \over 2}$& $-{1 \over 4}$& $-{1 \over 2}$& $1$ & $- {1 \over 4}$& $- {1 \over 2}$&$-{1 \over 2}$&$-1$\\
$\bar\chi^{({1 \over 2}, -{1 \over 2},  {3 \over 2})}$, $\bar\chi^{({1 \over 2},  {3 \over 2}, -{1 \over 2})}$& $\overline\lambda_3 Z_3$, $\overline\lambda_4 Z_2$& $-{1 \over 2}$& ${1 \over 4}$ & ${1 \over 2}$& $1$ & $-{1 \over 4}$& ${1 \over 2}$ &${1 \over 2}$&$-1$\\ \hline
\end{tabular}
\end{equation}
We note in passing that certain 5D spin-1/2 fields that mix with the gravitini do not satisfy all the rules derived from this dimensional reduction; they do satisfy the relation \eno{CRagain}, but while they have $|q_1| = 1/2$, they also have $|m_2| = |p_1| = 5/12$.   This indicates that those fields do not descend from a normal spin-1/2 kinetic term in 6D.  The most natural explanation is that they are actually modes of the 6D gravitino, polarized in the $z$-direction, though this has not been checked.

We have not yet obtained the coupling of the fermion to the gauge field $A_\mu$. One can consider a Pauli term in six dimensions,
\eqn{}{
{\cal L}_{\rm Pauli} = {i\over 2}  \sqrt{-\hat{g}} \bar\lambda^a \Gamma^{MNP} H_{MNP} \lambda_a \,,
}
and reduce this to five dimensions.  Consider the case where one of the indices is polarized in the $z$-direction.  (If the field strength is indeed self-dual, the other case will  contribute similarly.)  We then obtain schematically
\eqn{}{
{\cal L}_{\rm Pauli} \sim -{i \varepsilon\over 2}  e^{ \phi \over \sqrt{6}} \sqrt{-g} \bar\chi^a \gamma^{\mu\nu} F_{\mu\nu} \chi_a \,, 
}
which is the right scaling of the scalar for the 5D $A_\mu$ Pauli term.  Thus one can obtain $p_2$ simply by choosing the undetermined coefficient of the 6D Pauli term appropriately, since this is not fixed by anything else.

\subsection{Validity of the six-dimensional lift}

This lift, while valuable, does not include all information. In particular, the quantities $m_1$ and $q_2$ associated to each fermion, which are subleading near the horizon, do not arise. $m_1$ is the subleading term in the potential and seems to be invisible. There also does not seem to be any 6D coupling that generates $q_2$, the ordinary gauge coupling in 5D: there is no canonical coupling of a fermion to a 2-form field that reduces to this. (It does come from the reduction of a 6D one-form; but then one cannot produce the Pauli term, which is more important in the near-horizon limit.)

If one takes the 6D fermion or its 5D reduction and calculates the resulting Dirac equation, one gets a truncation of the equation analyzed in previous sections. In general, the first-order equations for $r  \to 0$ are 
\eqn{}{
\left( \partial_r + \left({a \over r^2} + b \right) \sigma_3 + \left( {c \omega \over r^2} + d \omega + f \right) i \sigma_2 + {P \over r} \sigma_1 \right) \psi = 0\,,
}
where the various constants are defined in \eno{Constants}; all these terms are required to obtain the correct limit of the Dirac equation. The 6D fermion, however, gives the equation with $b$, $d$ and $f$ dropped. While this might seem valid since they are all ${\cal O}(r^0)$ while the remaining terms are ${\cal O}(r^{-1})$ or higher, it is actually not consistent; in generating the second-order equations there   are cross terms $ab/r^2$ that are no smaller than $P^2/r^2$. Thus we are not entitled to keep $P$ if we drop $b$, $d$ and $f$. The ultimate source of this subtlety is that the two components of $\psi$ can be of different orders in the expansion in $r$. This is reflected in the fact that there is no rigorous scaling we can perform as $r$ gets small that lets us keep only $P$ as a subleading term.

The effect of $b$, $d$ and $f$ in our analysis is to provide a constant term containing $m_1$, $m_2$ and $q_2$ in the expression for $\nu$, which can play against the momentum-dependent $P$ term (see equation~\eno{NuSquared}.) We are thus only honestly entitled to neglect $b$, $d$ and $f$  when they are much smaller than $P$. In the limit of large $P$ we get for $\nu$,
\eqn{}{
\nu(|P|\gg1) \to |P| \mp {1 \over 2} \approx |P| \,.
}
Thus the six-dimensional lift is only fully self-consistent, with no additional terms required, in this limit. However, even though this limit does not in general obtain for the values of $k$ we examine, the lift is nonetheless useful for the resolution of the singularity, for the proof of the relations between $m_2$, $q_1$ and $p_1$, and for the intuition that excitations below the gap correspond to spacelike momenta. Subleading terms in $r$ presumably can be introduced to resolve these issues, at the cost of giving up the simple $AdS_3 \times \mathbb{R}^3$ form.

\section*{Acknowledgments}

We have enjoyed useful discussions and correspondence with Mike Hermele, Shamit Kachru, Elias Kiritsis, Hong Liu, Leo Radzihovsky, and Subir Sachdev.   The work of O.D.\ was supported by the Department of Energy under Grant No.~DE-FG02-91-ER-40672.  The work of S.S.G.\ was supported by the Department of Energy under Grant No.~DE-FG02-91ER40671. The work of C.R.\ was supported in part by European Union's Seventh Framework Programme under grant agreements (FP7-REGPOT-2012-2013-1) no 316165,
PIF-GA-2011-300984, the EU program ``Thales'' MIS 375734  and was also co-financed by the European Union (European Social Fund, ESF) and Greek national funds through the Operational Program ``Education and Lifelong Learning'' of the National Strategic Reference Framework (NSRF) under ``Funding of proposals that have received a positive evaluation in the 3rd and 4th Call of ERC Grant Schemes''.

\section*{Appendix A: A note on second-order decoupled Dirac equations}

In sections~\ref{GravitySec}-\ref{2QBHSec} we use the second-order equations \eno{DiracSecondOrder} associated to $U_\pm$, as described in \cite{Gubser:2012yb}. In our previous work \cite{DeWolfe:2012uv}, as well as the numerical solutions in section~\ref{Nequals4Sec}, we use $\psi_\pm$ directly, and the associated second-order equations,
\eqn{SecondOrderDirac}{
 \psi''_{ \pm} - F_\pm  \psi'_{ \pm}  + \left( \mp X' - X^2 + Y^2- Z^2 \pm X F_\pm \right)\psi_{ \pm}\,,
}
where
\eqn{FDef}{
F_\pm \equiv \partial_r \log \left(\mp Y+Z \right) \,.
}
The two formulations are rather similar, and both are correct. The main difference, other than the introduction of a few factors of $i$, is that the formulation in terms of $U_\pm$ involves $X$ and $Z$ in the coefficient of the first-order term \eno{pDef}, while the formulation in terms of $\psi_\pm$ has $Y$ and $Z$ instead \eno{FDef}. Because the energy $\omega$ is contained in $Y$, this distinction is relevant when taking a small-$\omega$ limit. The definition of $F_\pm$ puts $\omega$ in the denominator, and near-horizon expansions come with extra powers of $(r-r_H)/\omega$. Such terms cannot be ignored; in the inner region they are of order one, while in the outer region they actually diverge. This ruins the near-horizon expansion for (2+1)-charge black holes at $\omega \neq 0$. Thus it is preferred to work with the quantities $U_\pm$.
The analysis of \cite{DeWolfe:2012uv} remains correct, however; the inner region was solved using the first-order equations only, and was matched to \cite{Faulkner:2009wj}. In the outer region, $\omega = 0$ was taken strictly before the near-horizon limit, which gives correct expressions.

One may ask whether the $\psi_\pm$ formulation affects the 2-charge black hole. For that case one finds
\eqn{}{
F_\pm = - {2 \over r} \mp {P \over c \omega} + {\cal O}(r)\,, 
}
which implies that
\eqn{}{
\mp X' \pm X F_\pm = -{a P \over c \omega r^2} + {\cal O}\left(1 \over r\right)\,,
}
and one ends up with the equation $\eno{TwoQEqn}$ but with one term flipped:
\eqn{Flipped}{
{c \omega P \over a} \to {a P \over c \omega} \,.
}
Interestingly, this does not affect the near-horizon analysis. For solutions away from $|\omega| = \Delta$, this term is dropped as subleading. For solutions at $|\omega| = \Delta$, the two terms \eno{Flipped} are the same.

\section*{Appendix B: Five- and six-dimensional spinor conventions}

We use a mostly plus signature metric, but a ``mostly-minus"-type Clifford algebra, meaning that
\eqn{}{
\{ \gamma^{\underline{\mu}},  \gamma^{\underline{\nu}} \} = - 2 \eta^{\underline{\mu\nu}} \,,
}
where we use the underline to denote flat-space indices. Five-dimensional gamma matrices are $4 \times 4$,
and since we are in odd dimension there is a relation,
\eqn{}{
\gamma^{\underline{4}} = \gamma^{\underline{0}} \gamma^{\underline{1}} \gamma^{\underline{2}} \gamma^{\underline{3}} \,.
}
One can define conjugation/transpose matrices $B$ and $C$ obeying 
\eqn{}{
B \gamma^\mu B^{-1} = (\gamma^\mu)^* \,, \quad \quad C \gamma^\mu C^{-1} = (\gamma^\mu)^T \,.
}
and satisfying
\eqn{}{
B^{-1} = B \,, \quad  B^T = B^* = -B \,, \quad \quad C^{-1} = C^T = -C \,,  \quad C^* = C \,.
}
One choice is $B \equiv i \gamma^{\underline{4}}$, $C \equiv i \gamma^{\underline{0}} \gamma^{\underline{4}}$ which makes  $\gamma^{\underline{0}}$, $\gamma^{\underline{1}}$, $\gamma^{\underline{2}}$ and $\gamma^{\underline{3}}$ imaginary, while $\gamma^{\underline{4}}$ is real; $\gamma^{\underline{0}}$ and $\gamma^{\underline{4}}$ are antisymmetric while $\gamma^{\underline{1}}$, $\gamma^{\underline{2}}$ and $\gamma^{\underline{3}}$ are symmetric.  

Define $8 \times 8$ gamma matrices in six dimensions via
\eqn{}{
\Gamma^{\underline{\mu}} = \gamma^{\underline{\mu}} \otimes \sigma_1 \,, \quad \quad
\Gamma^{\underline{5}} = \mathbb{I} \otimes i\sigma_2\,,
}
and the six-dimensional chirality matrix is
\eqn{}{
\Gamma \equiv \Gamma^{\underline{0}} \Gamma^{\underline{1}} \Gamma^{\underline{2}} \Gamma^{\underline{3}} \Gamma^{\underline{4}} \Gamma^{\underline{5}} = \mathbb{I} \otimes \sigma_3 \,,
}
obeying as usual
\eqn{}{
\Gamma^2 = 1 \,,  \quad \{ \Gamma, \Gamma^\mu \} =0 \,.
}
 We also have 6D conjugation/transpose matrices,
\eqn{RealitySix}{
B_6 \Gamma^M B_6^{-1} = (\Gamma^M)^* \,, \quad \quad C_6 \Gamma^M C_6^{-1} = (\Gamma^M)^T \,,
}
defined as
\eqn{ReduceB}{
B_6 = B \otimes \mathbb{I}_{2 \times 2} \,, \quad \quad
C_6 = C \otimes \sigma_1 \,.
}
which in the particular basis mentioned above take the form $B_6 \equiv -i \Gamma^{\underline{4}}  \Gamma^{\underline{5}} \Gamma$, $C_6 =  \Gamma^0 B_6 = -i \Gamma^{\underline{0}} \Gamma^{\underline{4}} \Gamma^{\underline{5}} \Gamma$. Since $\sigma_1$ is real and symmetric, the $\Gamma^{\underline{\mu}}$ inherit the same reality/symmetry properties as the $\gamma^{\underline{\mu}}$.  Thus $\Gamma^{\underline{0}}$, $\Gamma^{\underline{1}}$, $\Gamma^{\underline{2}}$ and $\Gamma^{\underline{3}}$ are imaginary, while $\Gamma^{\underline{4}}$ and $\Gamma^{\underline{5}}$ are real, while $\Gamma^{\underline{0}}$, $\Gamma^{\underline{4}}$ and $\Gamma^{\underline{5}}$ are antisymmetric while $\Gamma^{\underline{1}}$, $\Gamma^{\underline{2}}$ and $\Gamma^{\underline{3}}$ are symmetric.   In addition, the chirality matrix $\Gamma$ is real and symmetric.

The fermionic reduction is complicated by the presence of symplectic Majorana spinors in five and six dimensions. 
The index $a, b, \ldots$ runs over multiple spinors, and these will be related in pairs by the symplectic Majorana condition.
We raise and lower symplectic Majorana spinor indices with the antisymmetric matrix $\Omega_{ab}$,
\eqn{}{
\chi_a =\Omega_{ab} \chi^b \,, \quad \quad
\chi^a = \Omega^{ab} \chi_b \,,
}
where $\Omega^{ab}$ is the inverse of $\Omega_{ab}$, $\Omega^{ab}\Omega_{bc} = \delta^a_c$; we may take conventions where $\Omega_{12} = -\Omega^{12} = 1$ for each pair of spinors.
 The symplectic Majorana condition is expressed in terms of the reality matrix $B$,
\eqn{SympMaj}{
\chi^a = B (\chi_a)^* = B \Omega_{ab} (\chi^b)^* \,.
}
Defining the barred spinor as
\eqn{}{
\bar\chi^a \equiv (\chi_a)^\dagger \gamma^{\underline{0}} \,,
}
we may also express the symplectic Majorana condition as
\eqn{SympMaj2}{
\bar\chi^a = (\chi^a)^T C\,.
}
We can use the latter form to prove the bilinear ``Majorana flip" expression for symplectic Majorana spinors,
\eqn{}{
\bar{\chi}^a \gamma^{\mu_1} \cdots \gamma^{\mu_n} \psi^b 
=\bar{\psi}^b \gamma^{\mu_n} \cdots \gamma^{\mu_1} \chi^a  \,.
}
This constrains possible terms in the Lagrangian for symplectic Majorana spinors; some terms require a relative sign between the two spinors of the pair in order to end up nonzero. 
It is useful to use a Pauli matrix notation for the symplectic Majorana space of each spinor pair to denote this.  If we agree that $\bar\chi$ stands for $\bar\chi^a$ and $\chi$ for $\chi_b$, the allowed terms are
\eqn{}{
{\cal L} = {1\over 2} \left(i \bar\chi \gamma^\mu \nabla_\mu \chi - m  \bar\chi \tau_3 \chi + q \bar\chi \gamma^\mu A_\mu \tau_3 \chi + ip \bar\chi \gamma^{\mu\nu} F_{\mu\nu} \chi \right) \,,
}
leading to the Dirac equation,
\eqn{}{
\left( i \gamma^\mu \nabla_\mu  - m \tau_3  + q \gamma^\mu A_\mu \tau_3  + ip  \gamma^{\mu\nu} F_{\mu\nu}  \right) \chi = 0 \,.
}
Here we use $\tau$ to stand for Pauli matrices in the symplectic Majorana space, to distinguish them from the $\sigma$ which are Pauli matrices in the Clifford algebra space.  The factor of $\tau_3$ in the mass and gauge coupling terms makes up for the antisymmetry of the spinor contraction.  The kinetic and Pauli terms do not need such a factor due to the minus signs from integrating the derivative by parts and from the antisymmetry of $F_{\mu\nu}$, respectively.\footnote{In the literature symplectic Majorana Lagrangians are typically expressed in terms of a non-diagonalized basis for the spinors, with $\tau_2$ instead of $\tau_3$.  We are concerned with matching to our already-diagonalized eigenvectors, so we use the form above.  The transformation from one basis to the other preserves the symplectic Majorana condition.}

Completely analogous expressions to \eno{SympMaj}, \eno{SympMaj2} hold for six-dimensional symplectic Majorana spinors. In that case we also have 
\eqn{}{
[B_6, \Gamma] =0\,,
}
which along with $\Gamma^* = \Gamma$ implies that the symplectic Majorana condition is compatible with a Weyl condition
\eqn{}{
 \Gamma \lambda = \pm \lambda \,,
}
allowing symplectic Majorana-Weyl spinors.
 In these conventions, a 6D Weyl spinor is simply an 8-component spinor with only the top four or bottom four components nonzero; if it is symplectic Majorana-Weyl, then thanks to \eno{ReduceB} the four nonzero components obey precisely the 5D symplectic Majorana condition.

\section*{Appendix C: Details of fermionic Kaluza-Klein reduction}

Here we provide the details of the reduction of the 6D Majorana-Weyl spinor to five dimensions discussed in section~\ref{6DSec}.
Given the ansatz for the decomposition of the 6D metric,
\eqn{}{
d\hat{s}^2 &= e^{2 \alpha \phi} ds^2 + e^{2 \beta \phi} (dz + {\cal A})^2
}
the sechsbeins are
\eqn{}{
\hat{e}^{\;\underline{\mu}}_{\;\;\;\nu} = e^{\alpha \phi} e^{\;\underline{\mu}}_{\;\;\;\nu} \,, \quad\quad
\hat{e}^{\;\underline{\mu}}_{\;\;\;z} = 0 \,, \quad\quad
\hat{e}^{\;\underline{z}}_{\;\;\;z} = e^{\beta\phi}\,, \quad\quad
\hat{e}^{\;\underline{z}}_{\;\;\;\nu} = e^{\beta\phi} {\cal A}_\nu\,,
}
where $e^{\;\underline{\mu}}_{\;\;\;\nu}$ is the 5D f\"unfbein.   The inverse sechsbeins are
\eqn{}{
\hat{e}^{\;\mu}_{\;\;\;\underline{\nu}} = e^{-\alpha \phi} e^{\;\mu}_{\;\;\;\underline{\nu}} \,, \quad \quad
\hat{e}^{\;\mu}_{\;\;\;\underline{z}} = 0 \,, \quad \quad
\hat{e}^{\;z}_{\;\;\;\underline{z}} =  e^{-\beta\phi}\,, \quad \quad
\hat{e}^{\;z}_{\;\;\;\underline{\nu}} = - e^{-\alpha \phi} e^{\;\mu}_{\;\;\;\underline{\nu}} \,{\cal A}_\mu \,.
}
The spin connection is then
\eqn{}{
\hat\omega_\mu^{\;\;\underline{\nu\rho}} &= \omega_\mu^{\;\;\underline{\nu\rho}} +\alpha\left( 
 e^{\;\underline{\nu}}_{\;\;\;\mu} \partial^{\underline{\rho}} \phi - 
 e^{\;\underline{\rho}}_{\;\;\;\mu} \partial^{\underline{\nu}} \phi \right)
- {1 \over 2} e^{2(\beta - \alpha)\phi}{\cal A}_\mu {\cal F}^{\underline{\nu\rho}}  \,, \cr
\hat\omega_\mu^{\;\;\underline{\nu z}} &=  {1 \over 2} e^{(\beta - \alpha)\phi} {\cal F}^{\;\;\;\underline{\nu}}_{\mu} - \beta e^{(\beta - \alpha) \phi}{\cal A}_\mu \partial^{\underline{\nu}} \phi \,, \cr
\hat\omega_z^{\;\;\underline{\nu\rho}} &=- {1 \over 2} e^{2(\beta - \alpha)\phi} \, {\cal F}^{\;\underline{\nu\rho}} \,, \cr
\hat\omega_z^{\;\;\underline{\nu z}} &= - \beta e^{(\beta - \alpha) \phi} \, \partial^{\underline{\nu}} \phi \,,
}
where all flat indices on the right-hand side are obtained from the curved-space indices with the 5D f\"unfbein.  We notice that due to the origin of the spin connection terms from $\hat{e}^{\underline{z}} = e^{\beta \phi} (dz + {\cal A})$, we can write the last terms of the first and second lines as
\eqn{SpinConRelations}{
\hat\omega_\mu^{\;\;\underline{\nu\rho}} &= \omega_\mu^{\;\;\underline{\nu\rho}} +\alpha\left( 
 e^{\;\underline{\nu}}_{\;\;\;\mu} \partial^{\underline{\rho}} \phi - 
 e^{\;\underline{\rho}}_{\;\;\;\mu} \partial^{\underline{\nu}} \phi \right)
+{\cal A}_\mu \hat\omega_z^{\;\;\underline{\nu\rho}} \cr
\hat\omega_\mu^{\;\;\underline{\nu z}} &=  {1 \over 2} e^{(\beta - \alpha)\phi} {\cal F}^{\;\;\;\underline{\nu}}_{\mu} +{\cal A}_\mu \hat\omega_z^{\;\;\underline{\nu z}}  \,.
}
This will be useful in what follows.

Consider the reduction of the 6D fermion kinetic term \eno{Kinetic6D} with the fermion ansatz \eno{FermionReductionAnsatz}. In all of our reductions below, we will end up with $\mathbb{I}_{2 \times 2}$ or $\sigma_3$ in the extra two-component space, and the 8-component inner product between spinors will collapse to a 4-component inner product; because the fermion bilinears will be diagonal in the two-component space, the factors of $e^{\pm i n \varphi_3}$ will cancel between the fermion and its conjugate.  Since $\Gamma = \mathbb{I}_{4 \times 4} \otimes \sigma_3$, the appearance of $\sigma_3$ will give us a factor of the chirality $\varepsilon$.

 First consider the terms involving the partial derivative.  There are three such terms:
\eqn{KineticDecomp}{
{\cal L}_{\partial} = \sqrt{- \hat{g}} {i\over2}(\lambda_a)^\dagger \Gamma^{\underline{0}} \left(\Gamma^{\underline{\mu}}\, \hat{e}^\nu_{\;\;\;\underline{\mu}}  \partial_\nu 
+ \Gamma^{\underline{5}} \hat{e}^z_{\;\;\underline{z}}  \partial_z +
\Gamma^{\underline{\mu}} \,\hat{e}^z_{\;\;\;\underline{\mu}} \partial_z 
\right)
\lambda_a \,, 
}
where no fourth term appears since $\hat{e}^\nu_{\;\;\underline{z}} = 0$.  These three terms will become the 5D kinetic term, mass term and KK gauge coupling term, respectively.  Consider first the kinetic term.
We have
\eqn{}{
\Gamma^{\underline{0}} \Gamma^{\underline{\mu}} = \gamma^{\underline{0}} \gamma^{\underline{\mu}} \otimes \mathbb{I}_{2 \times 2} \,,
}
and so the 8-component Weyl spinors reduce to four-component spinors as described above.  The $e^{\pm in \varphi_3}$ factors cancel.  We get one term where the partial derivative acts on the $\chi_a$,
\eqn{}{
{\cal L}_{\rm kin} = e^{\phi \over \sqrt{6}} \sqrt{- g} {i\over2} e^{2 \eta \phi \over \sqrt{6}} (\chi_a)^\dagger \gamma^{\underline{0}}  \gamma^{\underline{\mu}}  e^{- \phi \over 2 \sqrt{6}} e^\nu_{\;\;\underline{\mu}}\partial_\nu \chi_a + \ldots
= \sqrt{-g} e^{(1 + 4 \eta) \phi \over 2 \sqrt{6}} {i\over2} \bar\chi^a \gamma^\mu \partial_\mu \chi_a + \ldots\,.
}
Thus to get the canonical kinetic term, we need to pick $\eta = -1/4$.  Another term results when the derivative acts on the exponential before it cancels, so the total result coming from the first term in \eno{KineticDecomp} is
\eqn{FiveDKinetic}{
{\cal L}_{\rm kin} =  \sqrt{-g} \left( {i\over2} \bar\chi^a \gamma^\mu \partial_\mu \chi_a - {i \over 8 \sqrt{6}}
\bar\chi^a \gamma^\mu (\partial_\mu \phi) \chi_a \right)
\,.
}
The latter part will be canceled by the spin connection.

For the other two terms in \eno{KineticDecomp} we will need $\partial_z$ derivatives of the spinor,
\eqn{}{
\partial_z \lambda_1 = { in \over L} \lambda_1 \,, \quad \quad \partial_z \lambda_2 = -{in\over L} \lambda_2 \,,
}
or in Pauli matrix notation
\eqn{}{
\partial_z \lambda = {i n \over L} \tau_3 \lambda\,.
}
Using the gamma matrix identity
\eqn{}{
\Gamma^{\underline{0}} \Gamma^{\underline{5}}  =- \gamma^{\underline{0}}\,  \sigma_3 \,,
}
we obtain for the middle term in \eno{KineticDecomp},
\eqn{}{
{\cal L}_{\rm mass} &= e^{\phi \over \sqrt{6}} \sqrt{- g} {n \over2L} e^{-{ \phi \over 2 \sqrt{6}}} (\chi_a)^\dagger \sigma_3 \gamma^{\overline{0}} e^{3 \phi \over 2 \sqrt{6}}  (\tau_3)_a^{\;\;b} \chi_b  \,, \cr
&= e^{2 \phi \over \sqrt{6}} \sqrt{- g} {n\varepsilon\over2L} \bar\chi^a (\tau_3)_a^{\;\;b} \chi_b \,,
}
where the $\sigma_3$ got us a factor of the chirality $\varepsilon$.

Consider now the final term in \eno{KineticDecomp}.  This gives
\eqn{}{
{\cal L}_{\rm KK gauge} &= \sqrt{-\hat{g}} {i \over 2} (\lambda_a)^\dagger \Gamma^{\underline{0}} \Gamma^{\underline{\mu}} \,\hat{e}^z_{\;\;\;\underline{\mu}} \partial_z \lambda_a \,, \cr
&= \sqrt{- g} {n \over2L} \bar\chi^a \gamma^\mu {\cal A}_\mu
 (\tau_3)_a^{\;\;b} \chi_b \,,
 }
 or in terms of the 1Q gauge field $a_\mu = {\cal A}_\mu/2$,
 \eqn{}{
  {\cal L}_{\rm KK gauge}=   \sqrt{-g}{1 \over 2}\bar\chi \left( 2n \over L\right)  \gamma^\mu a_\mu
 \tau_3\chi \,.
}
Assembling these together, we get for the total terms  \eno{KineticDecomp} coming from the partial derivative in the 6D spinor kinetic term,
\eqn{TermsFromPartial}{
{\cal L}_{\partial} =  \sqrt{-g} \left( {i\over2} \bar\chi \gamma^\mu \partial_\mu \chi
  + {1 \over 2} \bar\chi \left( {n \varepsilon \over L} e^{2 \phi \over \sqrt{6}} \right) \tau_3 \chi
  + {1 \over 2}  \bar\chi \left( 2n \over L\right)  \gamma^\mu a_\mu
 \tau_3\chi    - {i \over 8 \sqrt{6}} \bar\chi \gamma^\mu (\partial_\mu \phi) \chi \right) \,.
}
Consider now the spin connection terms.  We will need to evaluate
\eqn{SpinConLag}{
{\cal L}_\omega = -  {i \over 8} \sqrt{-\hat{g}}\bar\lambda^a \Gamma^{\underline{M}} \,\hat{e}^N_{\;\;\;\underline{M}} \hat\omega_N^{\;\;\;\underline{PQ}} \Gamma_{\underline{PQ}} \lambda_a \,.
}
We have
\eqn{SpinDecomp}{
 \Gamma^{\underline{M}} \,\hat{e}^N_{\;\;\;\underline{M}} \hat\omega_N^{\;\;\;\underline{PQ}} \Gamma_{\underline{PQ}}  &=
  \Gamma^{\underline{\mu}} \,\hat{e}^\nu_{\;\;\;\underline{\mu}} \hat\omega_\nu^{\;\;\;\underline{PQ}} \Gamma_{\underline{PQ}} + 
  \Gamma^{\underline{\mu}} \,\hat{e}^z_{\;\;\;\underline{\mu}} \hat\omega_z^{\;\;\;\underline{PQ}} \Gamma_{\underline{PQ}} +
   \Gamma^{\underline{z}} \,\hat{e}^z_{\;\;\;\underline{z}} \hat\omega_z^{\;\;\;\underline{PQ}} \Gamma_{\underline{PQ}}  \cr
   &= \sigma_1 e^{-\alpha \phi} \gamma^\mu \left( \hat\omega_\mu^{\;\;\;\underline{PQ}} - {\cal A}_\mu  \hat\omega_z^{\;\;\;\underline{PQ}} \right) \Gamma_{\underline{PQ}} + i \sigma_2 e^{-\beta \phi} \hat\omega_z^{\;\;\;\underline{PQ}} \Gamma_{\underline{PQ}} \,.
}
Note that in the term in parentheses, the ${\cal A}_\mu  \hat\omega_z^{\;\;\underline{PQ}}$ piece entirely cancels against terms in the $\hat\omega_\mu^{\;\;\underline{PQ}}$ piece due to the relations \eno{SpinConRelations}.  The first term in \eno{SpinDecomp} then gives us
\eqn{}{
\sigma_1 e^{-\alpha \phi} \gamma^\mu \left( \hat\omega_\mu^{\;\;\;\underline{PQ}} - {\cal A}_\mu  \hat\omega_z^{\;\;\;\underline{PQ}} \right) \Gamma_{\underline{PQ}}  = 
\sigma_1 e^{-\alpha \phi} \left( \gamma^\mu \omega_\mu^{\;\;\;\underline{\nu\rho}} \gamma_{\underline{\nu\rho}} - 8 \alpha\gamma^\mu \partial_\mu \phi - \sigma_3 e^{(\beta -\alpha)\phi}\gamma^{\mu\nu} {\cal F}_{\mu\nu} \right) \,,
}
where to get the coefficient in the middle term we used the 5D Clifford identity
\eqn{}{
\gamma^\mu \gamma_{\mu\nu} = - 4 \gamma_\nu \,.
}
Meanwhile the second term in \eno{SpinDecomp} becomes
\eqn{}{
 i \sigma_2 e^{-\beta \phi} \hat\omega_z^{\;\;\;\underline{PQ}} \Gamma_{\underline{PQ}}  = - 2 \beta \sigma_1 e^{-\alpha \phi} \gamma^\mu \partial_\mu \phi - {i \over 2} \sigma_2 e^{(\beta - 2 \alpha)\phi} \gamma^{\mu\nu} {\cal F}_{\mu\nu} \,,
}
and combining these and using $\beta = - 3 \alpha$ we get
\eqn{}{
\Gamma^{\underline{M}} \,\hat{e}^N_{\;\;\;\underline{M}} \hat\omega_N^{\;\;\;\underline{PQ}} \Gamma_{\underline{PQ}}  = \sigma_1 e^{-\alpha \phi} \left( \gamma^\mu \omega_\mu^{\;\;\;\underline{\nu\rho}} \gamma_{\underline{\nu\rho}} - 2 \alpha\gamma^\mu \partial_\mu \phi - {1 \over 2} \sigma_3 e^{(\beta -\alpha)\phi}\gamma^{\mu\nu} {\cal F}_{\mu\nu} \right) \,.
}
Now plug this in to the reduction of the Lagrangian \eno{SpinConLag}.  The reduction of the metric determinant and the two $\lambda$'s into $\chi$'s give a factor $e^{\alpha \phi}$.  The $\Gamma^{\underline{0}}$ inside $\bar\lambda^a$ gives a factor $\sigma_1$.  We thus find the terms
\eqn{}{
{\cal L}_\omega =  \sqrt{-g} \bar\chi^a \left(- {i \over 8}  \gamma^\mu \omega_\mu^{\;\;\;\underline{\nu\rho}} \gamma_{\underline{\nu\rho}} +{i \over 8\sqrt{6}} \gamma^\mu \partial_\mu \phi+ {i\varepsilon \over 16} e^{(\beta -\alpha)\phi}\gamma^{\mu\nu} {\cal F}_{\mu\nu}  \right) \chi_a\,.
}
Finally we combine these with the rest of the terms we already found from the partial derivative pieces \eno{TermsFromPartial}.  The first new term simply gives the 5D spin connection, promoting the 5D partial derivative to a covariant derivative.  The second new term nicely cancels the extra $\gamma^\mu \partial_\mu \phi$ term generated by rescaling the spin-1/2 field to get a canonical kinetic term \eno{FiveDKinetic}.  The last term provides a Pauli coupling to the KK gauge field, which we rewrite using $a_\mu = {\cal A}_\mu/2$.  Our total 5D Lagrangian is now
\eqn{}{
e^{-1} {\cal L} =  {i\over2} \bar\chi \gamma^\mu \nabla_\mu \chi + {1 \over 2}\bar\chi \left(n \varepsilon \over L \right)e^{2 \phi \over \sqrt{6}} \tau_3 \chi + {1 \over 2} \bar\chi \left( 2 n \over L\right) \gamma^\mu a_\mu \tau_3 \chi 
+ {i \over 2} \bar\chi \left( \varepsilon \over 4 \right) e^{-2 \phi \over \sqrt{6}} \gamma^{\mu\nu} f_{\mu\nu} \chi  \,,
}
as given in \eno{FinalLagrangian}.

\bibliographystyle{JHEP}
\bibliography{Fermi2Q}

\end{document}